\documentclass[fleqn,usenatbib]{mnras}
\usepackage{natbib}
\usepackage{newtxtext,newtxmath}
\usepackage[T1]{fontenc}
\usepackage{ae,aecompl}

\usepackage{graphicx}	% Including figure files
\usepackage{amsmath}	% Advanced maths commands
\usepackage{amssymb}	% Extra maths symbols
\usepackage{float}
\usepackage{scrextend}
\usepackage{subcaption}
\usepackage[switch]{lineno}
\usepackage{booktabs}
\usepackage{longtable}
\usepackage[para,flushleft]{threeparttable}
\usepackage{multirow}
\makeatletter
\def\blfootnote{\xdef\@thefnmark{}\@footnotetext}
\makeatother

\title[1ES\,2344+514, intermittent EHBL]{An intermittent extreme BL Lac: MWL study of 1ES~2344+514 in an enhanced state}
\author[V.~A.~Acciari~et~al.]{\parbox{\textwidth}{\Large{
MAGIC Collaboration: V.~A.~Acciari$^{1}$,
S.~Ansoldi$^{2,24}$,
L.~A.~Antonelli$^{3}$,
A.~Arbet Engels$^{4\dagger\color{blue}\star}$,%\thanks{$^\color{blue}\star$ corresponding authors: M.~Manganaro, A.~Arbet Engels, D.~Dorner.~\href{mailto:contact.magic@mpp.mpg.de}{contact.magic@mpp.mpg.de}
A.~Babi\'c$^{6}$,
B.~Banerjee$^{7}$,
U.~Barres de Almeida$^{8}$,
J.~A.~Barrio$^{9}$,
J.~Becerra Gonz\'alez$^{1}$,
W.~Bednarek$^{10}$
% ...
}
\newline
\emph{\normalsize Affiliations are listed at the end of the paper}
}}

\date{Accepted XXX. Received YYY; in original form ZZZ}
\pubyear{2020}
\begin{document}
\label{firstpage}
\pagerange{\pageref{firstpage}--\pageref{lastpage}}
\maketitle
\clearpage
\begin{abstract}
Extreme High-frequency BL~Lacs (EHBL) feature their synchrotron peak of the broadband spectral energy distribution (SED) at \mbox{$\nu_{\rm s} \geq $10$^{17}$\,Hz}. The BL~Lac object 1ES~2344+514 was included in the EHBL family because of its impressive shift of the synchrotron peak in 1996. During the following years, the source appeared to be in a low state without showing any extreme behaviours. In August 2016, 1ES~2344+514 was detected with the ground-based $\gamma$-ray telescope FACT during a high $\gamma$-ray state, triggering multi-wavelength (MWL) observations.
We studied the MWL light curves of 1ES~2344+514 during the 2016 flaring state, using data from radio to VHE $\gamma$ rays taken with OVRO, KAIT, KVA, NOT, some telescopes of the GASP-WEBT collaboration at the Teide, Crimean, and St. Petersburg observatories, \textit{Swift}-UVOT, \textit{Swift}-XRT, \textit{Fermi}-LAT, FACT and MAGIC. With simultaneous observations of the flare, we built the broadband SED and studied it in the framework of a leptonic and an hadronic model.
The VHE $\gamma$-ray observations show a flux level of 55\% of the Crab Nebula flux above 300\,GeV, similar to the historical maximum of 1995. The combination of MAGIC and \textit{Fermi}-LAT spectra provides an unprecedented characterization of the inverse-Compton peak for this object during a flaring episode. The $\Gamma$ index of the intrinsic spectrum in the VHE $\gamma$-ray band is $2.04\pm0.12_{\rm stat}\pm0.15_{\rm sys}$. We find the source in an extreme state with a shift of the position of the synchrotron peak to frequencies above or equal to $10^{18}$\,Hz.\newline
\end{abstract}
%\begin{keywords}
%BL Lacertae objects: individual: 1ES~2344+514 -- galaxies: active -- gamma-%rays: galaxies
%\end{keywords}
%%%%%%%%%%%%%%%%% BODY OF PAPER %%%%%%%%%%%%%%%%%%

\section{Introduction}
\label{sec:intro}
Blazars\blfootnote{{$^{\color{blue}\star}$} corresponding authors: M.~Manganaro, A.~Arbet Engels, D.~Dorner.~\href{mailto:contact.magic@mpp.mpg.de}{contact.magic@mpp.mpg.de}
\newline {$\dagger$} also member of FACT Collaboration.
\newline {$\ddagger$} also member of MAGIC Collaboration.}  are radio-loud Active Galactic Nuclei (AGN), whose relativistic jets are aligned along our line of sight. A common classification of blazars into two main subcategories of BL~Lac objects (BL~Lacs, after the BL Lacertae object) and Flat Spectrum Radio Quasars (FSRQ) is based on the properties of their optical spectra \citep{1995PASP..107..803U}. BL~Lacs are generally characterized by their very weak or absent emission/absorption lines in the optical band. The majority of blazars emitting in the very-high-energy (VHE, $E > 100$\,GeV) band belong to the BL~Lacs family (57 so far)\footnote{http://tevcat.uchicago.edu/,\cite{2008ICRC....3.1341W}}.\par
Typically, BL~Lacs display a broadband spectral energy distribution (SED) characterized by a two-humped structure~\citep{2017MNRAS.469..255G}. The first hump of the SED, known as the synchrotron bump, is attributed to synchrotron radiation by relativistic electrons. In some cases, the host galaxy contributes to the first hump of the SED in the optical and infrared band, making a careful correction~\citep[see i.e.][]{2007A&A...475..199N} necessary. The high energy bump is often identified as being produced by inverse-Compton (IC) scattering of the synchrotron photons by the same population of electrons. This scenario represents the simplest leptonic model (one-zone Synchrotron Self Compton, SSC), but there are competing models including hadronic components~\citep[e.g.][]{2015MNRAS.448..910C}. The position of the synchrotron peak is widely used to further define three different types of BL~Lac objects: the low, intermediate, and high-frequency BL~Lac objects~\citep[see][and references therein]{1995ApJ...444..567P,2007Ap&SS.309...95B}. LBL (low-frequency BL~Lacs) have their synchrotron peak $\nu_{\rm s}$ in the sub-millimeter to infrared wavelengths \mbox{($\nu_{\rm s}<10^{14}$\,Hz)}, while HBL (high-frequency peaked BL~Lacs) in the ultraviolet (UV) to X-ray bands \mbox{($\nu_{\rm s}>10^{15}$\,Hz)}.  
IBL (intermediate-frequency BL~Lacs) feature their synchrotron peak in between the above mentioned ranges~\citep{1995ApJ...444..567P}.\par
In the past decades, new observations have revealed \citep{2007A&A...470..475A,2007A&A...473L..25A,2007A&A...475L...9A,2010ApJ...715L..49A} that a handful of sources show a $\nu_{\rm s}$ at unusually high X-ray energies with \mbox{$\nu_{\rm s} \geq $10$^{17}$\,Hz}. Based on this extreme behaviour, \citet{2001A&A...371..512C} proposed an additional category of BL~Lacs, the extreme high-frequency BL~Lacs (EHBL). As a consequence of an unusually high $\nu_{\rm s}$, EHBL can also have the IC hump peaking at unusually high frequency in the $\gamma$-ray band. 
Thus, the shift of the whole SED generally translates also in a particularly hard X-ray and VHE $\gamma$-ray spectra with a photon index \mbox{$\Gamma \lesssim 2$}.\par
Despite the above-mentioned spectral features characterizing EHBL, recent multi-wavelength (MWL) observations show that such objects can have very different temporal behaviour. Some of them, such as 1ES~0229+200~\citep{2007A&A...475L...9A} the archetypal EHBL, seem to constantly exhibit extreme properties. On the other hand, other objects have been identified to belong to the EHBL family only on a temporary basis \citep{2018A&A...620A.181A,2019MNRAS.486.1741F}. The TeV detected BL~Lac 1ES~2344+514 belongs to the latter group. So far, this object showed a $\nu_{\rm s}$ significantly above $10^{17}$\,Hz only during flaring state, as reported by \citet{2000MNRAS.317..743G}. \par
Located at a redshift of z=0.044 \citep{1996ApJS..104..251P}, 1ES~2344+514 was discovered by the \textit{Einstein} Slew Survey \citep{1992ApJS...80..257E} in the 0.2\,keV--4\,keV energy range.\par
The first detection in VHE $\gamma$ rays was obtained in 1995 by the Whipple 10\,m telescope during an intense flare, with a flux corresponding to $\sim 60$\% of the Crab Nebula flux above 350\,GeV \citep{1998ApJ...501..616C,2005ApJ...634..947S}. In 1996, 1ES~2344+514 showed a very variable behaviour in the X-ray band~\citep{2000MNRAS.317..743G} on a timescale of approximately 5\,ks when the source was at its brightest state: impressive rapid changes of the X-ray spectrum slope, together with a large shift by a factor of 30 or more of $\nu_{\rm s}$, put this source for the first time in the EHBL family. An analogous behaviour was observed a few months later in another source, Mrk~501, during an outburst in April 1997 \citep{1998ApJ...492L..17P}. In this case, the synchrotron peak shifted to energies around or above 100\,keV, making the source an additional member of the EHBL family. It should be noted that recently in 2012, Mrk~501 also exhibited an intermittent extreme behaviour during a low state \citep{2018A&A...620A.181A}.\par
Following the extreme event of 1ES~2344+514, MWL campaigns have been organized to study the source \citep{2007JPhG...34.1683G,2007ApJ...662..892A,2011ApJ...738..169A,2013A&A...556A..67A} and to model the broadband SED using simultaneous and quasi-simultaneous data. During most of these campaigns, the source was found in a lower state in the X-ray and VHE $\gamma$-ray band with respect to previous observations \citep{1998ApJ...501..616C,2000MNRAS.317..743G}, so the broadband SED obtained were mainly describing the source during low activity. In all those occasions, a one-zone SSC model was found to well describe the data. Different observations by IACT (Imaging Atmospheric Cherenkov Telescopes) over the past years have revealed the source to have variable flux states in the VHE $\gamma$-ray band. The integral flux is generally less than 10\% of the Crab Nebula flux, excluding two short flares with 60\% and 50\% of the Crab Nebula flux level \citep{1998ApJ...501..616C,2011ApJ...738..169A}. More recent VHE $\gamma$-ray data for this source have been presented in \citet{2017MNRAS.471.2117A}, where the temporal properties of 1ES~2344+514 are studied on short and long timescales in the VHE $\gamma$-ray band, and no significant flaring activity was observed since 2008.\par
In the present work, we report on the observations of a VHE $\gamma$-ray flare of 1ES~2344+514 in August 2016, detected by FACT (First G-APD Cherenkov Telescope), and followed up by many instruments, including the MAGIC (Major Atmospheric Gamma Imaging Cherenkov) telescopes, the \textit{Fermi}-LAT (Large Area Telescope), \textit{Swift}-XRT and \textit{Swift}-UVOT, TCS (Telescope Carlos S\'{a}nchez), KAIT (Katzman Automatic Imaging Telescope), KVA (Kungliga Vetenskapsakademien), Stella, LX-200, AZT-8, NOT (Nordic Optical Telescope) IAC80 and OVRO (Owens Valley Radio Observatory). We collected a dataset from simultaneous and quasi-simultaneous MWL observations of a flaring state. FACT and MAGIC are both IACT devoted to the study of VHE $\gamma$ rays. 
It is worth to note that, for this source, the combination of \textit{Fermi}-LAT and MAGIC data for the first time offers an unprecedented characterisation of the IC peak during a flaring state.\par
The paper is structured as follows. In Sect.~\ref{sec:obs}, the details of the observations performed by the instruments involved are reported, together with the description of the dedicated analysis. In Sect.~\ref{sec:MWL}, the MWL light curves and their variability are discussed. Sect.~\ref{sec:spectral} is devoted to the analysis of the spectra in the VHE $\gamma$-ray and X-ray band, to describe the IC and the synchrotron peak respectively. In Sect.~\ref{sec:SED}, the broadband SED is presented together with the modeling, while in Sect.~\ref{sec:extreme}, we discuss in detail the extreme behaviour of the source during this particular flaring state. Conclusions are drawn in Sect.~\ref{sec:summary}. 

\section{Multi-wavelength Observations}
\label{sec:obs}
In this section, the details of the observations and the data analysis for the various instruments are reported.
\subsection{VHE $\gamma$-ray observations with FACT and MAGIC}
\label{sec:VHE_obs}
The FACT telescope, located at the Observatorio del Roque de los Muchachos (ORM) in La Palma, has been observing at TeV energies since October 2011 \citep{2013JInst...8P6008A}. 
The excellent performance and stability of the used semi-conductor photosensors \citep{2014JInst...9P0012B} combined with the observing strategy maximizes the observation time and minimizes the observational gaps \citep{2019ICRC...36..665D}. An automatic on-site quick-look analysis provides with low latency publicly available results\footnote{\url{https://fact-project.org/monitoring}} and allows for 
More details on the analysis can be found in Appendix~\ref{sec:appendix}.\par
1ES~2344+514 has been monitored by FACT since August 2012 for a total of more than 1950\,hours (status at October 2019). When the flux found in the quick-look analysis exceeds 50\% of the flux of the Crab Nebula at TeV energies, an alert is issued to MWL partners. For 1ES~2344+514, seven flare alerts have been issued in five years. One of these alerts was sent on MJD~57610 (10 August 2016) and triggered the MWL campaign presented here. 
The dataset used in this study includes 118.6\,hours after data-quality selection from 65~nights between MJD~57568 (29 June 2016) and MJD~57645 (14 September 2016). To remove data obtained during bad weather, the cosmic-ray rate \citep{2017ICRC...35..779H} has been used after correcting it for the effect of zenith distance and threshold, as described in \citet{2017ICRC...35..612M} and \citet{2019APh...111...72B}. 
The light curve has been determined using the analysis chain described in \citet{2019ICRC...36..630B} calculating the excess rate using the {\it Lightcurve Cut}. 
Based on the excess rate of the Crab Nebula, the standard candle at TeV energies, the dependencies of the excess rate from trigger threshold (which changes with the ambient light conditions) and zenith distance are determined and corrected for. Also the correction for the effect of the Saharan Air Layer (SAL) is applied (details in Appendix~\ref{sec:appendix}). Using simulated data, the energy threshold of this analysis is determined to be 775\,GeV for a Crab-Nebula-like spectrum. For a harder spectrum, as measured for 1ES\,2344+514 at VHE $\gamma$ rays during previous observing campaigns, the energy threshold is accordingly higher \citep[$\sim810$\,GeV for a spectral slope of 2.46 as in][]{2017MNRAS.471.2117A}.\par
MAGIC is a stereoscopic system consisting of two 17\,m diameter IACT located at the ORM, on the Canary Island of La Palma. The current sensitivity for observations at small angular distances from the zenith \mbox{(zd: $15^\circ< \mathrm{zd} < 30^\circ$)} above 289\,GeV is \mbox{$(0.72\pm0.04)\%$} of the Crab Nebula flux in 50\,h \citep{2016APh....72...76A}.\par
MAGIC started to observe 1ES~2344+514 on MJD~57611 (11 August 2016), triggered by the enhanced activity in the VHE range revealed by FACT. 
We collected data in the zenith distance range of \mbox{$23^\circ< \mathrm{zd} < 33^\circ$} and the analysis was performed using the standard MAGIC analysis framework MARS \citep{zanin2013,2016APh....72...76A}.  After the applied quality cuts, the surviving events amounted to \mbox{N$_{\rm on}= 533$; N$_{\rm off} = 256 \pm 7.6$; N$_{\rm ex} = 277.0 \pm 24.3$} in a total of 0.87~hr of data after quality cuts. A full description of the MAGIC systematic uncertainties can be found in \citet{2016APh....72...76A} and references therein. The source was detected with a significance of $13\sigma$ in 0.62\,hr. The flux above 300\,GeV was \mbox{$(7.2 \pm 0.9)\times 10^{-11}\,\textrm{cm}^{-2}\,\textrm{s}^{-1}$},  which corresponds to the 55\% of the Crab Nebula flux in the same energy range.\par
The following night, MJD~57612 (12 August 2016), MAGIC observed again the source for 0.48\,hr, and found a more than three times lower flux with respect to the previous night, corresponding to \mbox{$(2.1 \pm 0.4)\times 10^{-11}\,\textrm{cm}^{-2}\,\textrm{s}^{-1}$} above 300\,GeV (16\% of the Crab Nebula flux above 300\,GeV). The significance for MJD~57612 (12 August 2016) was found to be $4\sigma$ in 0.48\,hr of observations.\par

\subsection{HE $\gamma$-ray observations with \textit{Fermi}-LAT}
The LAT is a pair-conversion telescope on board the \textit{Fermi} satellite and has been monitoring the high-energy (HE, \mbox{0.1\,GeV $< E < 100$\,GeV)} $\gamma$-ray sky for almost 12\,yr. The instrument is able to cover a wide energy range from 20\,MeV to $>300$\,GeV \citep{2009ApJ...697.1071A,2012ApJS..203....4A}. The LAT normally operates in survey mode with an all-sky coverage on a $\sim 3$\,hr timescale. The analysis presented here was carried out using the unbinned-likelihood tools from the version \texttt{v10r0p5} of the \textit{Fermi} Science Tools software\footnote{http://fermi.gsfc.nasa.gov/ssc/}.\par
For this work, we considered a region of interest (ROI) with a radius of 15$^{\circ}$ around 1ES~2344+514 and selected \texttt{Source} class events in an energy range from 0.1\,GeV to 300\,GeV. In addition, we applied a cut of 52$^{\circ}$ for the rocking angle as well as a maximum zenith distance of 100$^{\circ}$ to reduce contamination from the Earth limb photons. We used the instrument response function \texttt{P8R2\_SOURCE\_V6} and the diffuse background models\footnote{http://fermi.gsfc.nasa.gov/ssc/data/access/lat/\\BackgroundModels.html} \texttt{gll\_iem\_v06} and \texttt{iso\_P8R2\_SOURCE\_V6\_v06}.\par
A first unbinned-likelihood analysis was performed considering LAT data over a 6-month period between MJD~57509 (01 May 2016) and MJD~57691 (30 October 2016). All point sources from the 3FGL catalog \citep{2015ApJS..218...23A} within 15$^{\circ}$ from 1ES~2344+514 were included in the model. During the fit, the flux normalisations and the spectral parameters of the sources were free to vary, but were fixed to the catalog values for the sources which are located further than 6$^{\circ}$ from 1ES~2344+514 or have a detection significance in the 3FGL catalog lower than $5\sigma$. As in the 3FGL catalog, we modeled 1ES~2344+514 with a simple power law with index $\Gamma$. After this first fit, sources resulting in a test statistics TS $< 10$~\citep{1996ApJ...461..396M} were removed from the model and a second unbinned-likelihood analysis was performed considering the simplified model. We also searched for new sources potentially detected by the LAT within 6$^{\circ}$ from the target. A TS sky map does not reveal significant sources not included in the 3FGL or in the 4FGL catalogs. In addition, we note that none of sources reported in the 4FGL catalog, but absent from the 3FGL catalog, were significantly detected.\\
The output model from the second unbinned-likelihood analysis was used to build a light curve between MJD~57567.5 (28 June 2016) and MJD~57644.5 (13 September 2016) in the  0.3\,GeV--300\,GeV energy range. For the light curve calculation, the spectral shape of 1ES~2344+514 and the normalisation of the diffuse background models were left free to vary. All the remaining sources had their spectral shapes parameters fixed to the values obtained from the second unbinned-likelihood analysis. Since 1ES~2344+514 is a faint source for LAT, we adopted a 7 days binning \citep{2009ApJ...707.1310A}. 
For time bins resulting in a detection with a TS below 4, we quote an upper limit at 95\% confidence level (C.L.).\par 
Over the period covered by the light curve, we report a clear detection with TS~$=110.5$. The spectrum is best described with a power-law index of $\Gamma=1.9\pm0.1$. To build the broadband SED used for the modeling, we consider a spectrum averaged over 1 month centred around the MAGIC observations (MJD 57596.5 to MJD 57626.5). The best-fit spectral index is in this case $\Gamma=1.7\pm0.2$ (TS~$=41$).

\subsection{X-ray observations with \textit{Swift}-XRT}
The \textit{Neil Gehrels Swift observatory (Swift)} has pointed to the source eight times from August 2016 to November 2016. The raw images by the X-ray Telescope \citep[\textit{XRT},][]{2004SPIE.5165..201B} onboard the \textit{Swift} satellite, are analyzed. These eight observations, performed in photon counting mode for the period from MJD~57613.52 (13 August 2016) to MJD~57696.18 (4 November 2016), have a total exposure time of $\sim4.1$\,hr with an average integration time of 0.44\,hr each. 
Following the procedure described by \citet{2017A&A...608A..68F} and assuming fixed equivalent Galactic hydrogen column density  $N_\textrm{H} = 1.5 \times 10^{21}\,\rm cm^{-2}$ \citep{2005A&A...440..775K}, we fitted the spectrum of each observation assuming all possible combination of pixel-clipping together with two mathematical models (power law and log parabola), and normalization energy $E_0=0.3$\,keV.
During the MAGIC campaign, the five X-ray spectra during the MWL campaign are quite hard (photon index, $\Gamma_{\rm XRT} \lesssim 2.1$) and in all cases they can be described by a power-law model. The constant flux ($F_{2-10\,\mathrm{keV}}$) hypothesis is rejected at more than $8~\sigma$ C.L..

\subsection{UV observations with \textit{Swift}-UVOT}

The UV data of 1ES~2344+514 were obtained by the \textit{Swift}-UVOT telescope in three UV bands~\citep[W1, M2 and W2,][]{2005SSRv..120...95R}.
The aperture photometry analysis was performed using standard Swift/UVOT software tools available within the HEASOFT package (version 6.24) along with calibration data from the latest release of CALDB (version \texttt{20170922}). An aperture radius of 5\,arcsec was used for all the filters. The background flux level was estimated in a circle of 20\,arcsec radius located close to 1ES~2344+514. Both background and source regions were verified not to be contaminated with light from any nearby objects. The fluxes were dereddened following the Equation (2) from \citet{2009ApJ...690..163R} using the value of $E(B - V) = 0.1819$ \citep{2011ApJ...737..103S}. In the UV band, the contribution of the host galaxy was negligible.

\subsection{Optical and near-infrared observations with Tuorla, WEBT and KAIT}
The Tuorla blazar monitoring program\footnote{\url{http://users.utu.fi/kani/1m/}} obtained optical ($R$ band, Cousins) data of the source between MJD~57500 (22 April 2016) and MJD~57650 (19 September 2016). The Kungliga Vetenskapsakademien (KVA) telescope and Nordic Optical Telescope (NOT), both located at the ORM, were used for these observations. The data are analyzed using the differential photometry method described by \citet{2018A&A...620A.185N}.\par
Additional optical and near-infrared data have been acquired thanks to the Whole Earth Blazar Telescope\footnote{\url{http://www.oato.inaf.it/blazars/webt/}} \citep[WEBT, e.g.][]{2007A&A...464L...5V,2017Natur.552..374R} consortium. The WEBT was born in 1997 with the aim of organising monitoring campaigns on specific blazars in a MWL context. In 2007, the WEBT started the GLAST-AGILE Support Program \citep[GASP, e.g.][]{2009A&A...504L...9V}, to provide low-energy data of a list of selected objects that could complement the high-energy observations by the $\gamma$-ray satellites. 1ES~2344+514 is one of the sources regularly monitored within the GASP.
In this framework, observations were performed at the Crimean (AZT-8), St.\ Petersburg (LX-200), and Teide (IAC80, STELLA and TCS telescopes) observatories.\par
From the observed $R$-band flux densities of the source we subtracted  the contribution by the host galaxy and nearby companions according to the prescriptions by \citet{2007A&A...475..199N}. This contribution depends on both the aperture radius adopted for the photometry and the seeing. We then corrected for the Galactic extinction $A_R=0.458$ from \citep{2011ApJ...737..103S}, as it was done with UVOT data sample. 
Intercalibration among the different datasets was refined by checking the consistency of the R-band light curve.\par
%KAIT
Optical images of 1ES~2344+514 were obtained with the Katzman Automatic Imaging Telescope \citep[KAIT;][]{2001ASPC..246..121F} at the Lick Observatory.  All images were reduced using a custom pipeline \citep{2010ApJS..190..418G} before doing the photometry. We applied a 9-pixel aperture (corresponding to $7.2''$) for photometry.
Several nearby stars were chosen from the Pan-STARRS1\footnote{\url{http://archive.stsci.edu/panstarrs/search.php}} catalog for calibration, their magnitudes were transformed into the Landolt magnitudes using the empirical prescription presented by \citet{2012ApJ...750...99T}, equation 6. All the KAIT images were taken without filter (namely in $clear$ band), which is the closest to the $R$ filter \citep[see][]{2012ApJ...750...99T}. We therefore calibrate all the $clear$ band result to the Pan-STARRS1 $R$-band magnitude.
Data from KAIT have been as well corrected for the host-galaxy contribution together with the other optical data, with the procedure described above.\par
In the near-infrared (NIR), data were obtained at the Teide Observatory with the TCS telescope in the framework of the GASP-WEBT collaboration. To estimate the contribution of the host galaxy to the JHK photometry, we proceeded as done e.g.\ in \citet{2010A&A...524A..43R} for the host galaxy of BL Lacertae. We used a de Vaucouleurs galaxy profile with an effective radius of 10.9\,arcsec \citep{2007A&A...475..199N} to estimate the host-galaxy contribution within the aperture radius used (10\,arcsec). This resulted in 48\% of the total host flux. We corrected the observed magnitude of the host, $R=13.90$ \citep{2007A&A...475..199N} for the Galactic extinction and applied the \citet{2001MNRAS.326..745M} colour indices for elliptical galaxies to estimate the deabsorbed magnitudes of the host in $JHK$ bands. We converted them into absorbed flux densities and subtracted 48\% of these values from the observed source fluxes. Finally, we corrected for the Galactic extinction values from \citep{2011ApJ...737..103S} to get the deabsorbed jet fluxes.

\subsection{Radio observations with OVRO}
The 15\,GHz data of 1ES~2344+51.4 were obtained within the OVRO 40-m Telescope blazar monitoring program \citep{2011ApJS..194...29R}. The OVRO 40\,m uses off-axis dual-beam optics and a cryogenic pseudo-correlation receiver with a 15~GHz centre frequency and 3\,GHz bandwidth. Calibration is achieved using a temperature-stable diode noise source to remove receiver gain drifts and the flux density scale is derived from observations of 3C~286 assuming the \citet{1977A&A....61...99B} value of 3.44\,Jy at 15.0\,GHz. The systematic uncertainty of about 5\% in the flux density scale is not included in the error bars.  Complete details of the reduction and calibration procedure can be found in \citet{2011ApJS..194...29R}.

\section{Analysis of the MWL light curves}

\label{sec:MWL}
\begin{figure*}
	\centering
	\includegraphics[width=2.1\columnwidth]{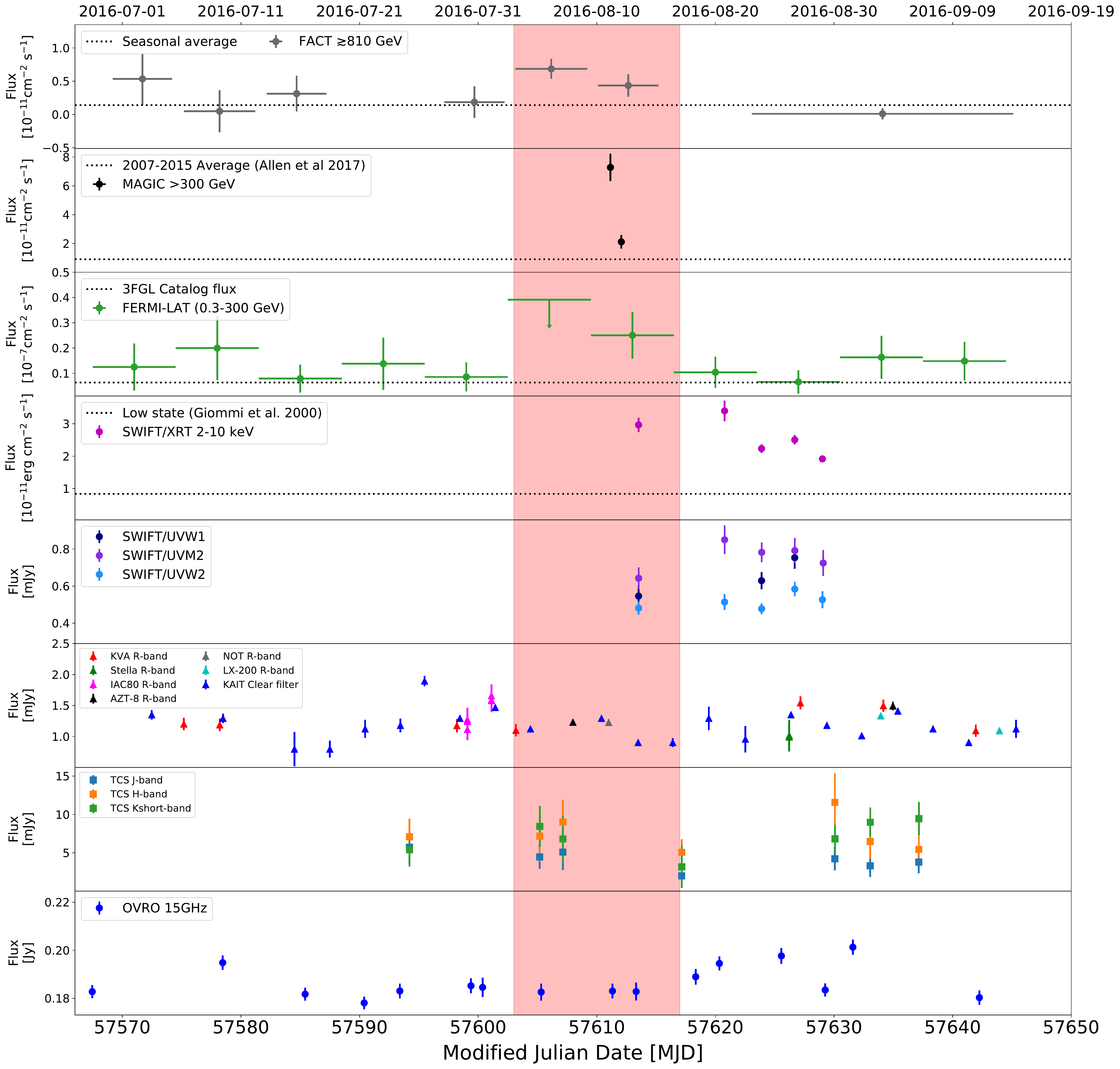}
    \caption{MWL light curve of 1ES~2344+514 from MJD~57567 (28 June 2016) to MJD~57645 (14 September 2016). The observations were carried out using (from top to bottom) FACT (energy threshold $\sim 810$\,GeV), MAGIC ($>300$\,GeV, \textit{Fermi}-LAT (0.3\,GeV--300\,GeV), \textit{Swift}-XRT (2\,keV--10\,keV), \textit{Swift}-UVOT (W1, M2 and W2 filter), KVA, NOT, IAC80 and Stella at Teide, AZT-8 at Crimean Observatory, LX-200 at St. Petersburg ($R$ band), KAIT (optical, clear filter), TCS at Teide ($J$, $H$, $K$ filters), and OVRO (15\,GHz). 
    In the \textit{Fermi}-LAT light curve we quote an upper limit at 95\% C.L. for time bins having TS~$<4$. The shadowed band represents the VHE $\gamma$-ray activity window. The band is centred on MJD~57610 (10 August 2016), which is the night in which FACT sent the alert which triggered MWL observations. The night of MJD~57613 (13 August 2016) is the night chosen for building the broadband SED. HE and VHE $\gamma$-ray panels show the flux in number of photons, while the other panels show the flux density in the respective energy bands. Optical and NIR data from 1ES~2344+514 shown here are corrected for the host galaxy contribution and galactic reddening.}
    \label{fig:MWL_lc}
\end{figure*}
The MWL light curves from radio to VHE energies are shown in Fig.~\ref{fig:MWL_lc} and include all observations from MJD~57567 to MJD~57645 (20 June 2016 to 14 September 2016). The MAGIC, \textit{Swift}-XRT, \textit{Swift}-UVOT, optical \& OVRO light curves are daily binned. The \textit{Fermi}-LAT and FACT light curves are both 7-day binned. As the VHE flux observed by FACT after MJD~57620 (20 August 2016) is consistent with no signal and showing no significant variability, the last FACT time bin is integrated over a period of $\sim 1$ month. For comparison purposes, we show as dashed black lines the seasonal average (from June 2016 to December 2016) in the FACT light curve, and the 7-years average taken from \citet{2011ApJ...738..169A} in the MAGIC light curve. Regarding  \textit{Fermi}-LAT and \textit{Swift}-XRT observations, we select as reference values the flux from the 3FGL catalog~\citep{2015ApJS..218...23A} and the low flux measured by \citet{2000MNRAS.317..743G} in 1998, respectively.\par
The MAGIC observations were triggered thanks to the FACT detection of an enhanced state on MJD~57610 (10 August 2016). On MJD~57611 (11 August 2016), the MAGIC measurements indeed show a strong flaring episode in the VHE $\gamma$-ray band, which corresponds to \mbox{$F(>300$\,GeV) $= (7.2 \pm 0.9) \times 10^{-11}$\,cm$^{-2}$\,s$^{-1}$}.
In order to directly compare with the Whipple observations, we compute the flux for the 11 August 2016 (MJD 57611) also above 350\,GeV and obtain \mbox{$F(>350$\,GeV) $= (5.8 \pm 0.8) \times 10^{-11}$\,cm$^{-2}$\,s$^{-1}$}, corresponding to $0.56\pm0.08$ of the Crab Nebula flux, which is comparable to the historical maximum.\par 
In the following night (MJD~57612--12 August 2016), a strong decrease is visible with \mbox{$F(>300$\,GeV) $= (2.10 \pm 0.46) \times 10^{-11}$\,cm$^{-2}$\,s$^{-1}$}, corresponding to $0.16\pm0.04$ of the Crab Nebula flux. Consequently, we observe a reduction of the flux by a factor $\sim 3.4$.
It therefore constitutes a clear indication of a day timescale variability, which was already reported in \citet{2011ApJ...738..169A} as well as during the 1995 historical flare. So far, no significant intra-night variability in the VHE $\gamma$-ray band has been detected for 1ES~2344+514. Significant variability on such short timescale is also not found during any of the MAGIC and FACT observations. The most recent long-term study of VHE $\gamma$-ray emission from 1ES~2344+514 was published by the VERITAS collaboration \citep{2017MNRAS.471.2117A}. Despite a flux variability detected from seasons to season, the observations from 2008 to 2015 showed no significant flare. Over this period of approximately 7 years, the averaged flux is $\sim 0.04$ of the Crab Nebula flux. When assuming this state as the emission baseline, the MAGIC observations performed on MJD~57611 (11 August 2016) reveal a $\sim 14$-fold higher flux.\par 
The FACT long-term light curve also displays an enhanced flux level of \mbox{$\sim 0.5 \times 10^{-11}$\,cm$^{-2}$\,s$^{-1}$} above 810\,GeV between MJD~57603 (03 August 2016) and MJD~57615 (15 August 2016). This corresponds to \mbox{$\sim 0.2$} of the Crab Nebula flux, which is above the seasonal average ($\sim 0.05$ of the Crab Nebula; dashed line in the top panel of Fig.~\ref{fig:MWL_lc}). The last time bin, averaged over about one month, shows that the source enters again a low state after MJD~57620 (20 August 2016). Over the latter period, the source is not significantly detected by FACT. The measured flux lies below the seasonal average. The computation of an upper-limit at 95\% C.L. results in \mbox{$0.17\times 10^{-11}$\,cm$^{-2}$\,s$^{-1}$} \mbox{($\sim 0.05$} of the Crab Nebula flux). \par
In this work, we present for the first time simultaneous HE and VHE $\gamma$-ray observations during a flaring state of 1ES~2344+514. The ${Fermi}$-LAT 0.3\,GeV-300\,GeV light curve is shown in the third panel from the top in Fig.~\ref{fig:MWL_lc}. The firm detection (TS = 110.5) obtained between MJD~57567.5 (28 June 2016) and MJD~57644.5 (13 September 2016) allows an unprecedented constrain of the IC bump of the SED. The highest weekly averaged flux seen by LAT during this campaign is \mbox{$F(0.3$--300\,GeV) $= (2.5 \pm 0.9) \times 10^{-8}$\,cm$^{-2}$\,s$^{-1}$} and is temporally coincident with the VHE high state observed by MAGIC and FACT. Nevertheless, the relatively large statistical uncertainty prevents to claim a significant HE flux increase, which could be associated to the VHE flare. We do not find any sign of short-term variability based on a 2-days binning light curve close to the VHE flare. Between MJD~57602.5 (2 August 2016) and MJD~57609.5 (9 August 2016) the FACT light curve is at its maximum and reveals an enhanced activity over several days in the VHE $\gamma$-ray band. In the same time period, LAT observations only results in a TS below 3 and an upper limit at 95\% C.L. is quoted. We stress that the low TS is mainly due to a small exposure that is about 14 times lower than for the other time bins, and may not be caused by a drop in the GeV flux.
Between MJD~57567.5 (28 June 2016) and MJD~57644.5 (13 September 2016), the averaged flux yields \mbox{$F(0.3$--300\,GeV) $= (1.2 \pm 0.2) \times 10^{-8}$\,cm$^{-2}$\,s$^{-1}$}. This amounts to a flux that is around \mbox{$\sim 2$} times higher than what is reported in the 3FGL catalog and confirms an enhanced state during the overall studied period. The best fit spectral parameters using a power-law model result in a spectrum with a photon index of \mbox{$\Gamma$ = $1.9\pm0.1$}.\par
The VHE $\gamma$-ray flaring episode is also accompanied with an elevated X-ray emission state. The first $\textit{Swift}$-XRT observation took place on MJD~57613 (13 August 2016). The $\textit{Swift}$-XRT daily binned light curve shows an energy flux of \mbox{$\sim 3 \times 10^{-11}$\,erg\,cm$^{-2}$\,s$^{-1}$} close to the simultaneous MAGIC-FACT observations. In comparison with the average flux obtained from the multi-year $\textit{Swift}$-XRT light curve shown in \citet{2013A&A...556A..67A}, this yields an approximately three times higher energy flux. However, we note that such emission state remains moderately high compared to the flare that happened in December 2007, where a peak flux of \mbox{$F_{2--10\,\mathrm{keV}}$ $= (6.28 \pm 0.31) \times 10^{-11}$\,erg\,cm$^{-2}$\,s$^{-1}$} was detected \citep{2011ApJ...738..169A}. Unfortunately, no strictly simultaneous X-ray observations are available at the highest VHE $\gamma$-ray state seen by MAGIC on MJD~57611 (11 August 2016). Thus, the possibility of a higher X-ray flux during the latter day compared to MJD~57613 (13 August 2016) remains.\par
The light curve from \textit{Swift}-UVOT is shown in the fourth panel from the bottom. The points are simultaneous with the  X-ray light curve, and they show a hint of activity after the VHE $\gamma$-ray flare, a short period of time which unfortunately is not covered by the other instruments. \par
We collected optical data acquired by several instruments in the $R$ band. The best sampled curve was obtained with the KAIT telescope. 
The highest flux density in the optical band is registered on MJD~57595 (26 June 2016), which is few days before the highest VHE $\gamma$-ray flux observed by MAGIC, MJD~57611 (11 August 2016). It corresponds to 1.89\,mJy which is not particularly high for this source \citep[see e.g.][]{2007ApJ...662..892A}.\par
The maximum flux density in the radio band is observed on MJD~57631 (31 August 2016) and corresponds to 0.201\,Jy. The latter value is slightly higher than the averaged one ($\sim 0.16$\,Jy) recorded in \citet{2013A&A...556A..67A}, making the radio light curve interesting: it would have been important to complement the radio data with a VLBI (Very Long Baseline Interferometry) map, which can identify possible emitted knots or reconnections in the jet, but unfortunately there are none available which could be considered for our study. 1ES~2344+514 is very faint usually for VLBI observations.\par
As the optical ($R$ band) and OVRO data are better sampled with respect to the other wavebands, we use the Discrete Correlation Function~\citep{1988ApJ...333..646E} to search for correlations with potential time lag. No significant correlation between these two bands was found.

\subsection{Variability}
Based on the data shown in Fig.~\ref{fig:MWL_lc}, we carry out a search for flux variability in the different energy bands.\par
As mentioned in the previous section, VHE variability has been observed from yearly to daily timescale for this object. The strong flux decrease between the two MAGIC observations constitutes an additional clear evidence of variability at a timescale of $\sim 1$\,day. Unfortunately, because of the limited amount of available data, more sophisticated variability analysis based on MAGIC data is not possible. The much larger dataset collected by FACT offers the possibility to search for variability on a longer timescale. A fit of a constant flux to the weekly binned light curve yields a \mbox{$\chi^{2}/\text{d.o.f.}= 18.4/6$}, based on which the hypothesis of a constant emission can be rejected at a $\sim 3\sigma$ level.\par 
A variability index of $\sim 100$ in the 3FGL catalog  indicates that 1ES~2344+514 is unlikely (<1\%) a steady HE emitter \citep{2015ApJS..218...23A}. Compared to other typical TeV blazars such as Mrk~421 ($\sim 190$) or Mrk~501 ($\sim 250$) this value is rather low, but remains significantly higher than other established EHBL like 1ES~0229+200 ($\sim 50$) or 1ES~2037+521 ($\sim 40$). A constant fit to the $\textit{Fermi}$-LAT light curve shown in Fig.~\ref{fig:MWL_lc} gives a \mbox{$\chi^2$/d.o.f. = $4.94/9$}, which is consistent with a constant flux emission in the 0.3\,GeV-300\,GeV range. The low variability in HE is a well known feature of EHBL, but it could be due to the long integration time (because of the low luminosity) that washes out the flux variations.\par
From the $\textit{Swift}$-XRT light curve, a general trend of a decreasing X-ray flux is visible from  \mbox{$\sim 3 \times 10^{-11}$\,erg\,cm$^{-2}$\,s$^{-1}$} to \mbox{$\sim 2 \times 10^{-11}$\,erg\,cm$^{-2}$\,s$^{-1}$} along the days after the flare. A constant fit to the five available observations gives a \mbox{$\chi ^2/\text{d.o.f.}= 35.6/4$}, which matches a p-value of \mbox{$3.5\times 10^{-7}$}, and is therefore a $\sim 5\sigma$ significance detection of flux variability. X-ray flux variability is a typical behaviour of the source and was reported several times for flaring episodes as well as during particularly low state. The most notable and strongest X-ray variability is described in \citet{2000MNRAS.317..743G} during the December 1996 flare, when for the first time roughly hourly variability is clearly visible. During low state, X-ray variability is reported in \citet{2013A&A...556A..67A}, though on longer timescales ($\sim 1$/day). Here also, only variability on roughly daily timescale can be claimed and no shorter timescale variability is detected. The X-ray spectral variability will be discussed in details in Sect.~\ref{sec:spectral}.\par
Regarding the lower energy bands, in the UV and near-IR no significant variability is detected. On the other hand, the optical $R$-band and 15\,GHz radio light curves are statistically inconsistent with a constant flux and from the fit we obtain \mbox{$\chi^{2}/\text{d.o.f.}= 460.3/41$} and \mbox{$\chi^{2}/\text{d.o.f.}= 74.3/15$}, respectively. This reveals some evolution also at the lowest energy of the broadband emission spectrum.\par

\begin{figure}
    \centering
    \includegraphics[width=1\linewidth]{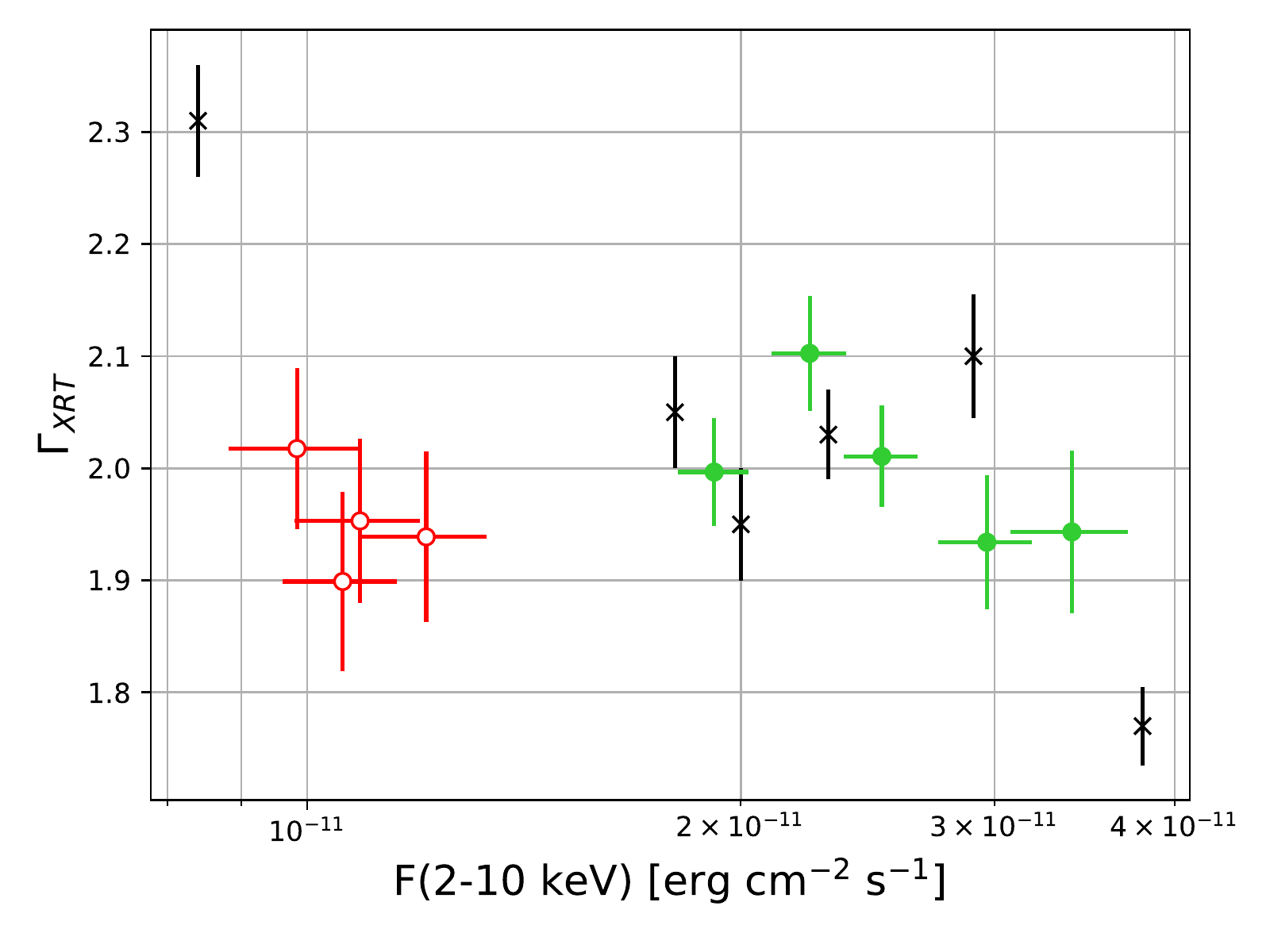}
    \caption{$\Gamma_{\rm XRT}$ vs.~2\,keV--10\,keV flux. Solid green data points corresponds to the observations up to MJD~57629--29 August 2016 (high state), while open red data points matches the observations from MJD~57688--27 October 2016 (low state). Black asterisks correspond to the archival data from \textit{BEPPO-SAX} observations between 1996 and 1998 (\citealt{2000MNRAS.317..743G}).}
    \label{fig:x_ray_flux_vs_index}
\end{figure}

\begin{table*}
\caption{Results of the spectral fits for X-ray data (power law).} 
    \centering
    \begin{tabular}{c|c|c|c|c|c}
    \hline
Date &  Exposure & $F_{2-10\,\text{keV}}$ &  $\Gamma_{\rm XRT}$  &  $\chi^{2}$/d.o.f.    \\ %header
    &    (s)    & $10^{-11}$\,erg\,cm$^2$ &    &    \\ %header units
\hline
\hline
13 August 2016  (MJD~57613) & 1651  & 2.96$\pm$0.22 & 1.93$\pm$0.06  & 36/36 \\
\hline
20 August 2016  (MJD~57620) & 1134  & 3.39$\pm$0.32 & 1.94$\pm$0.07  & 24.2/26 \\
\hline
23 August 2016  (MJD~57623) & 1833  & 2.23$\pm$0.13 & 2.10$\pm$0.05  & 33.1/43 \\
\hline
26 August 2016  (MJD~57626) & 2333  & 2.51$\pm$0.15 & 2.01$\pm$0.05  & 44.24/56 \\
\hline
29 August 2016  (MJD~57629) & 1987  & 1.92$\pm$0.11 & 2.00$\pm$0.05  & 45.39/51 \\
\hline
27 October 2016 (MJD~57688) & 1456  & 1.06$\pm$0.10 & 1.90$\pm$0.08  & 24.4/20 \\
\hline
29 October 2016 (MJD~57690) & 1453  & 0.98$\pm$0.10 & 2.02$\pm$0.07  & 22.4/26 \\
\hline
01 November 2016 (MJD~57693) & 1359  & 1.09$\pm$0.11 & 1.95$\pm$0.07  & 20.8/25 \\
\hline
04 November  2016 (MJD~57696) & 1518  & 1.21$\pm$0.12 & 1.94$\pm$0.08  & 19.5/25 \\
\hline
\hline
    \end{tabular}
    \label{tab:x_ray}
\end{table*}

\section{Spectral analysis}
\label{sec:spectral}
\subsection{X-ray spectral analysis and synchrotron peak identification}
\label{sec:spectral_Xray}
We study the X-ray emission by considering all \textit{Swift}-XRT observations of 1ES~2344+514 from MJD~57613 (13 August 2016) to MJD~57696 (4 November 2016). This represents a broader time range than the one presented in Fig.~\ref{fig:MWL_lc} and also provides a more comprehensive range of flux states. Table~\ref{tab:x_ray} summarizes the flux values together with the corresponding photon indices. The power-law index $\Gamma_{\rm XRT}$ versus the 2\,keV--10\,keV flux is plotted in Fig.~\ref{fig:x_ray_flux_vs_index}. A clear separation is visible between high and low flux states: the solid green symbols correspond to the five observations temporally closer to the flaring state in the VHE $\gamma$-ray band. Data after MJD~57698 (29 August 2016) are plotted with open red symbols. For these days, the 2\,keV--10\,keV flux is lower, \mbox{$\sim 10^{-11}$\,erg\,cm$^{-2}$\,s$^{-1}$}, which is comparable to the archival low state from MJD~50990 (26 June 1998), plotted with a black asterisk in Fig.~\ref{fig:x_ray_flux_vs_index}. Such a flux level in the 2\,keV--10\,keV band is also typical during low state in this energy band \citep{2011ApJ...738..169A}.\par
All spectra are well fitted with photon indices around or below 2 on the 0.3\,keV--10\,keV range. This hardness is typical of an EHBL \citep{2001A&A...371..512C}, thus in agreement with a location of $\nu_{\rm s}$ close or above 10$^{17}$\,Hz. Until now, all studies of 1ES~2344+514 have revealed the usual harder-when-brighter behaviour in the X-ray band. It has been reported during high \citep{2011ApJ...738..169A} as well as during low state \citep{2013A&A...556A..67A}. From Fig.~\ref{fig:x_ray_flux_vs_index}, this trend is however not visible. \textit{Swift}-XRT observations are fully consistent with a constant photon index and no spectral variability is visible. Based on a constant fit, the spectra are in agreement with a constant hard photon index of 2, at least during the considered period. This result contrasts with the previous observed strong spectral variability. Interestingly, if we exclude the strong X-ray flare of December 2007 \citep{2011ApJ...738..169A}, the dynamical range of the 2\,keV--10\,keV fluxes presented in this paper is quite typical for 1ES~2344+514. We are therefore probing typical X-ray states where one would naturally expect the standard harder-when-brighter trend that was reported in all previous studies. A comparison with the black asterisks in Fig.~\ref{fig:x_ray_flux_vs_index} that represents archival \textit{BEPPO-SAX} observations \citep{2000MNRAS.317..743G} clearly illustrates the peculiar behaviour of 1ES~2344+514 during 2016.\par
The closest spectrum in time to the VHE flare, from MJD~57613 (13 August 2016), has a photon index \mbox{$\Gamma = 1.93 \pm 0.06$} with no indication of curvature or steepening at higher energy. This constitutes a strong hint that we are describing the rising flank of the synchrotron component, and thus $\nu_{\rm s}$ is located at the edge, or beyond, the energy range covered by \textit{Swift}-XRT (i.e., $\geq 10^{18}$\,Hz) for this day. Based on archival data, \citet{2018A&A...620A.185N} estimated a $\nu_{\rm s}$ significantly lower, at $2.2 \times 10^{16}$\,Hz. Regarding the spectra obtained in October and November 2016, they show a similar hardness, despite the almost three times lower flux. As a comparison, for a similar 2\,keV--10\,keV flux, \citet{2011ApJ...738..169A} and \citet{2000MNRAS.317..743G} measured a much softer photon index of \mbox{$\Gamma \approx 2.3$--2.4}, clearly implying a peak located $\lesssim10^{17}$\,Hz.\par
\begin{figure}
    \centering
    \includegraphics[width=1\linewidth]{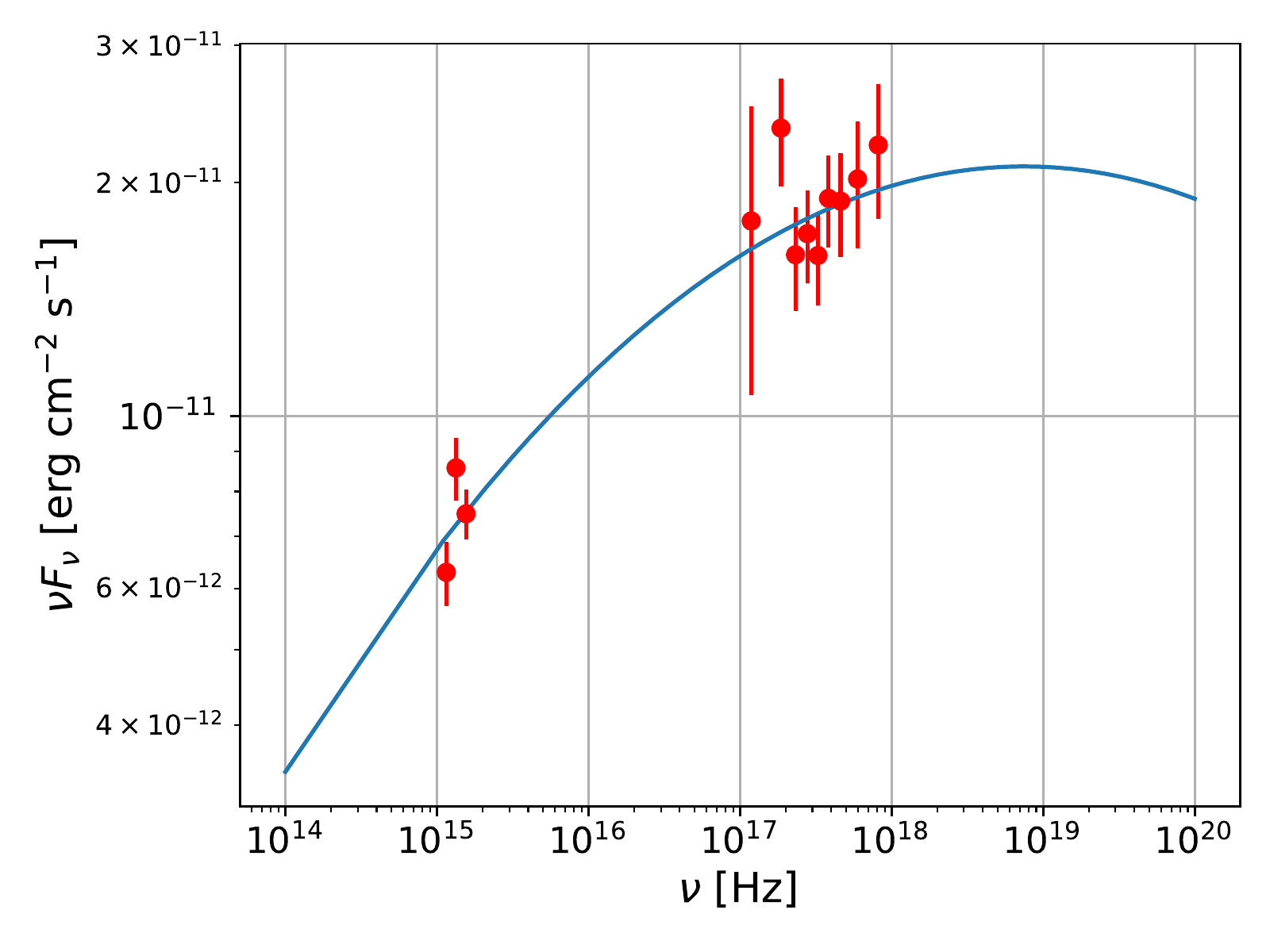} 
    \caption{Strictly simultaneous SED (MJD~57613 -- 13 August 2016) from \textit{Swift}-XRT, \textit{Swift}-UVOT is fitted with a log parabola in order to estimate the position of the synchrotron peak. The tentative value found is \mbox{$\nu_{\rm s} \approx 7.5 \times 10^{18}$\,Hz} \mbox{($\sim 30$ keV}--\mbox{$\chi^{2}/\mathrm{d.o.f.}= 8.97/9$)}. A more precise value deducted from the full broadband SED is presented in Sect.~\ref{sec:leptonic}.}
    
    \label{fig:synch_peak}
\end{figure}
None of the \textit{Swift}-XRT observation shows a statistical preference for a log-parabola shape. In order to search for spectral curvatures, which would help to constrain $\nu_{\rm s}$, we increase the statistics by grouping the observations according to their flux states and perform a spectral analysis on the summed datasets. We define three groups as following: the high flux state close to the VHE flare from MJD~57613 (13 August 2016) until MJD~57620 (20 August 2016), the intermediate flux state from MJD~57623 (23 August 2016) until MJD~57629 (29 August 2016) and the low flux state from MJD~57688 (27 October 2016) until MJD~57696 (04 November 2016). We find that only the intermediate flux state group shows a preference for a log-parabola shape. The derived $\nu_{\rm s}$ is $1.3\pm0.3$\,keV ($\sim3 \cdot 10^{17}$ Hz), in agreement with an extreme state. Regarding the high flux state group, no curvature is found and the spectral analysis on this summed dataset reveals a power-law index of 1.93$\pm$0.05 (\mbox{$\chi^{2}/\mathrm{d.o.f.}= 71.16/62$)}. This hardness supports a shift of the synchrotron peak above or at the edge of the \textit{Swift}-XRT passband (i.e., $\geq 10^{18}$\,Hz). The low state group gives a power-law index of 2.03$\pm$0.05 (\mbox{$\chi^{2}/\mathrm{d.o.f.}= 33.68/49$)}, indicating a peak around a few $10^{17}$\,Hz.\par
In an attempt to better locate and quantify the potential frequency shift of the peak close to the flare seen by FACT and MAGIC, we combine the SED from \textit{Swift}-XRT observations of MJD~57613 (13 August 2016) with strictly simultaneous UV data obtained with the \textit{Swift}-UVOT instrument (UVW1, UVM2, UVW2). The resulting plot is shown in Fig.~\ref{fig:synch_peak}. As in \citet{2013A&A...556A..67A}, we follow the same procedure presented in \citet{2007A&A...467..501T} and we fit of a log-parabola shape to the combined SED: 
\begin{equation}
    \nu F (\nu) = f_{0} \cdot 10^{-b \cdot (\log_{10}(\nu / \nu_{\rm s}))^{2}} \rm{erg\,cm}^{-2}s^{-1},
\end{equation}
where $\nu_{\rm s}$ is the peak location and $b$ the curvature. The resulting curve is shown in Fig.~\ref{fig:synch_peak}. The lack of data above $10^{18}$\,Hz leads to large uncertainties in the fitted parameters. The best fit value for the peak location is \mbox{$\nu_{\rm s} \sim7.5 \cdot 10^{18}$ Hz (${\sim}30$ keV)}. The same study presented in \citet{2013A&A...556A..67A} during rather low X-ray state \mbox{(F(2\,--10\,keV) $\approx 1 \times 10^{-11}$\,erg\,cm$^{-2}$\,s$^{-1}$)} yielded a peak between $0.3$\,keV and $3$\,keV, which is close to an order of magnitude lower. Despite being not able to precisely constrain the peak from the observations alone, Fig.~\ref{fig:synch_peak} and the measured hard photon index of \mbox{$\Gamma \approx 1.9$} confirm a shift above \mbox{$10^{18}$\,Hz} on MJD~57613 (13 August 2016). Interestingly, from the observations by \textit{BEPPO-SAX} during the 1996 flare \citep{2000MNRAS.317..743G} the 2\,keV--10\,keV flux level and spectral properties are comparable to our results on MJD~57613 (13 August 2016).\par
A further estimation of the $\nu_{\rm s}$ is given by means of the leptonic modeling in Sec.~\ref{sec:leptonic}.
\begin{figure*}
    \centering
    \includegraphics[width=2\columnwidth]{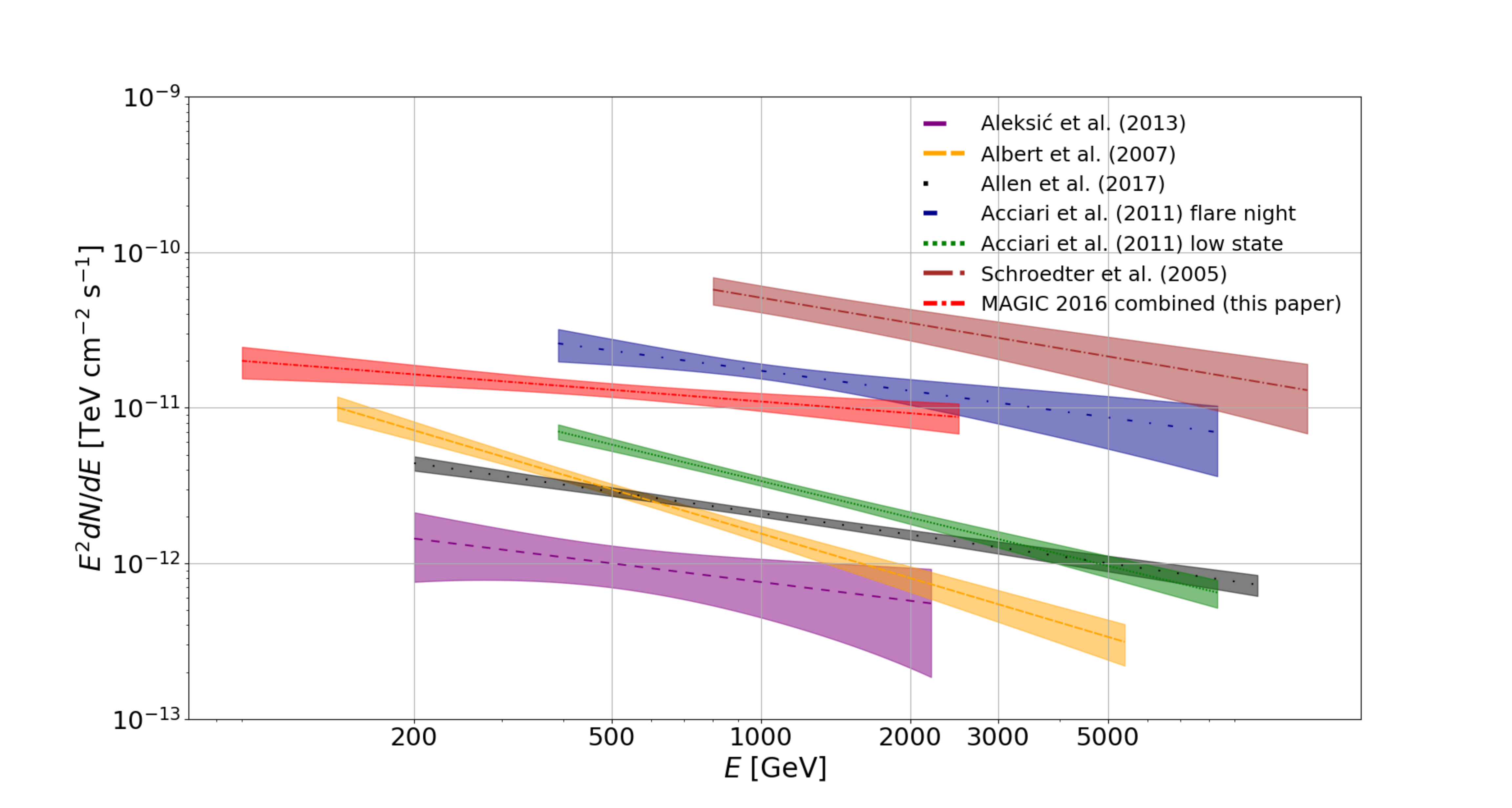} 
    \caption{Observed SED from all measurements of 1ES~2344+514 in the VHE $\gamma$-ray band. The respective parameters are listed in Table~\ref{table:gamma_previous}.}
    
    \label{fig:all_spectra}
\end{figure*}

\begin{table*}
\begin{threeparttable}
\caption{$\Gamma$ photon indices of the power-law fit and integral flux as percentage of the Crab Nebula C.N.$^a$ }   
\label{table:gamma_previous}      
\centering          
\begin{tabular}{ l c c c c c }   
\hline\hline 
 & Epoch$^b$ & $\Gamma$ (observed spectrum) &$\Gamma$ (intrinsic spectrum)& $F$ [\% C.N.] & $> E$  \\ %table heading
\hline\hline
Whipple & 1995$^{1}$  & 2.54 $\pm 0.17_{\rm stat} \pm 0.07_{\rm sys}$  & not reported & 63\%  & 350\,GeV   \\ 
\hline
MAGIC & 2007$^{2}$ & 2.95 $\pm 0.12_{\rm stat} \pm 0.2_{\rm sys}$ & 2.66 $\pm 0.50_{\rm stat} \pm 0.2_{\rm sys}$ & 10\% & 200\,GeV \\
\hline
\multirow{2}{*}{VERITAS} & \multicolumn{1}{l}{2007$-$2008$^{3}$  (low state)} & \multicolumn{1}{c}{2.78 $\pm 0.09_{\rm stat} \pm 0.15_{\rm sys}$}  & \multicolumn{1}{c}{$\sim 2.5$}  &  \multicolumn{1}{c}{7.6\%} & \multicolumn{1}{c}{300\,GeV} \\\cline{2-6}
\multirow{2}{*}{} & \multicolumn{1}{l}{2007$-$2008$^{3}$ (flare)} & \multicolumn{1}{c}{2.43 $\pm 0.22_{\rm stat} \pm 0.2_{\rm sys}$ }  & \multicolumn{1}{c}{ $\sim 2.1$}  &  \multicolumn{1}{c}{48\%} & \multicolumn{1}{c}{300\,GeV} \\\cline{1-6}
MAGIC  & 2008$^{4}$ & 2.4 $\pm 0.4_{\rm stat} \pm 0.2_{\rm sys}$ & 2.2 $\pm 0.4_{\rm stat} \pm 0.2_{\rm sys}$ & 2.5\% & 300\,GeV  \\
\hline  
VERITAS & 2007-2015$^{5}$ &  2.46 $\pm 0.06_{\rm stat} \pm 0.2_{\rm sys}$ & 2.15 $\pm 0.06_{\rm stat} \pm 0.2_{\rm sys}$ & 7\% & 350\,GeV  \\ 
\hline
MAGIC & 2016$^{6}$ & 2.25 $\pm 0.12_{\rm stat} \pm 0.15_{\rm sys}$&  2.04 $\pm 0.12_{\rm stat} \pm 0.15_{\rm sys}$ & 33\% &  300\,GeV \\ 
\hline
\hline
\end{tabular}
\begin{tablenotes}      
\small
$^a${Values determined above the energy reported in the last column from previous VHE $\gamma$-ray observations and from the most recent VHE $\gamma$-ray flare in 2016.}\\
$^b${Datasets described in detail: $^{1}$\cite{2005ApJ...634..947S}, $^{2}$\cite{2007ApJ...662..892A}, $^{3}$\cite{2011ApJ...738..169A},$^{4}$\cite{2013A&A...556A..67A}, $^{5}$\cite{2017MNRAS.471.2117A}, $^{6}$this work.}
\end{tablenotes}
\end{threeparttable}
\end{table*}

\begin{table*}
\caption{Parameters of the power-law fit to the VHE $\gamma$-ray spectra observed by MAGIC (this work).}             
\label{table:VHEspectrum}      
\centering          
\begin{tabular}{ l c c c c c c }     % 5 columns 
\hline\hline 
 MJD &  & $f_{0}$ [cm$^{-2}$\,s$^{-1}$\,TeV$^{-1}$] & $\Gamma$  & $\chi^{2}$/d.o.f. & Prob. \\  % table heading       
\hline\hline   
\multirow{2}{*}{combined} & \multicolumn{1}{l}{observed} & \multicolumn{1}{l}{$5.22\pm0.53_{\rm stat}\pm0.57_{\rm sys} \times 10^{-11}$} & \multicolumn{1}{l}{$2.25\pm0.12_{\rm stat}\pm0.15_{\rm sys}$} & \multicolumn{1}{l}{$10.51/8$}  & \multicolumn{1}{l}{23\%} \\\cline{2-6} % parameters and band from forward folding
\multirow{2}{*}{} & \multicolumn{1}{l}{intrinsic} & \multicolumn{1}{l}{ $6.60\pm0.67_{\rm stat}\pm0.72_{\rm sys} \times 10^{-11}$} & \multicolumn{1}{l}{$2.04\pm0.12_{\rm stat}\pm0.15_{\rm sys}$} & \multicolumn{1}{l}{$9.81/8$} & \multicolumn{1}{l}{28\%} \\\cline{1-6} % % parameters and band from forward folding
\multirow{2}{*}{57611} & \multicolumn{1}{l}{observed} & \multicolumn{1}{l}{$9.14\pm1.10_{\rm stat}\pm1_{\rm sys} \times 10^{-11}$} & \multicolumn{1}{l}{$2.33\pm0.15_{\rm stat}\pm0.15_{\rm sys}$} & \multicolumn{1}{l}{$4.10/8$}  & \multicolumn{1}{l}{84.8\%} \\\cline{2-6} % parameters and band from forward folding
\multirow{2}{*}{} & \multicolumn{1}{l}{intrinsic} & \multicolumn{1}{l}{ $1.14\pm0.14_{\rm stat}\pm0.12_{\rm sys} \times 10^{-10}$} & \multicolumn{1}{l}{$2.12\pm0.16_{\rm stat}\pm0.15_{\rm sys}$} & \multicolumn{1}{l}{$4.17/8$} & \multicolumn{1}{l}{84.1\%} \\\cline{1-6} % % parameters and band from forward folding
\multirow{2}{*}{57612} & \multicolumn{1}{l}{observed} & \multicolumn{1}{l}{$3.06\pm0.72_{\rm stat}\pm0.45_{\rm sys} \times 10^{-11}$} & \multicolumn{1}{l}{$2.22\pm0.41_{\rm stat}\pm0.15_{\rm sys}$} & \multicolumn{1}{l}{$5.87/8$}  & \multicolumn{1}{l}{66\%} \\\cline{2-6} % parameters and band from forward folding
\multirow{2}{*}{} & \multicolumn{1}{l}{intrinsic} & \multicolumn{1}{l}{ $3.44\pm0.87_{\rm stat}\pm0.51_{\rm sys} \times 10^{-11}$} & \multicolumn{1}{l}{$2.00\pm0.29_{\rm stat}\pm0.15_{\rm sys}$} & \multicolumn{1}{l}{$6.06/8$} & \multicolumn{1}{l}{64\%} \\\cline{1-6} % % parameters and band from forward folding
\end{tabular}
\end{table*}

\begin{table*}
\caption{Parameters of the log-parabola fit to the VHE $\gamma$-ray spectra observed by MAGIC (this work).}             
\label{table:VHEspectrumlogparabola}      
\centering          
\begin{tabular}{ l c c c c c c c}    
\hline\hline 
 MJD &  & $f_{0}$ [cm$^{-2}$\,s$^{-1}$\,TeV$^{-1}$] & $\alpha$ & $\beta$ & $\chi^{2}$/d.o.f. & Prob. \\  % table heading       
\hline\hline   
\multirow{2}{*}{combined} & \multicolumn{1}{l}{observed} & \multicolumn{1}{l}{$7.03\pm1.37_{\rm stat} \times 10^{-11}$} & \multicolumn{1}{l}{$2.38\pm0.22_{\rm stat}$} & \multicolumn{1}{l}{$1.08\pm0.79_{\rm stat}$} & \multicolumn{1}{l}{$8.90/7$}  & \multicolumn{1}{l}{26\%} \\\cline{2-7} % parameters and band from forward folding
\multirow{2}{*}{} & \multicolumn{1}{l}{intrinsic} & \multicolumn{1}{l}{ $8.41\pm1.63_{\rm stat} \times 10^{-11}$} & \multicolumn{1}{l}{$2.16\pm0.21_{\rm stat}$} & \multicolumn{1}{l}{$0.92\pm0.77_{\rm stat}$} &
\multicolumn{1}{l}{$8.72/7$} & \multicolumn{1}{l}{27\%} \\\cline{1-7} % % parameters and band from forward folding
\multirow{2}{*}{57611} & \multicolumn{1}{l}{observed} & \multicolumn{1}{l}{$1.08\pm0.21_{\rm stat} \times 10^{-10}$} & \multicolumn{1}{l}{$2.43\pm0.27_{\rm stat}$} & \multicolumn{1}{l}{$1.16\pm0.84_{\rm stat}$} &
\multicolumn{1}{l}{$5.56/7$}  & \multicolumn{1}{l}{58\%} \\\cline{2-7} % parameters and band from forward folding
\multirow{2}{*}{} & \multicolumn{1}{l}{intrinsic} & \multicolumn{1}{l}{ $1.30\pm0.26_{\rm stat} \times 10^{-10}$} & \multicolumn{1}{l}{$2.21\pm0.27_{\rm stat}$} & \multicolumn{1}{l}{$1.01\pm0.82_{\rm stat}$} &
\multicolumn{1}{l}{$5.47/7$} & \multicolumn{1}{l}{60\%} \\\cline{1-7} % % parameters and band from forward folding
\multirow{2}{*}{57612} & \multicolumn{1}{l}{observed} & \multicolumn{1}{l}{$3.70\pm2.03_{\rm stat} \times 10^{-11}$} & \multicolumn{1}{l}{$2.39\pm0.60_{\rm stat}$} & \multicolumn{1}{l}{$1.22\pm2.01_{\rm stat}$} &
\multicolumn{1}{l}{$4.56/7$}  & \multicolumn{1}{l}{71\%} \\\cline{2-7} % parameters and band from forward folding
\multirow{2}{*}{} & \multicolumn{1}{l}{intrinsic} & \multicolumn{1}{l}{ $4.40\pm2.39_{\rm stat} \times 10^{-11}$} & \multicolumn{1}{l}{$2.17\pm0.59_{\rm stat}$} & \multicolumn{1}{l}{$1.02\pm2.31_{\rm stat}$} &
\multicolumn{1}{l}{$4.56/7$} & \multicolumn{1}{l}{71\%} \\\cline{1-7} % % parameters and band from forward folding
\end{tabular}
\end{table*}

\subsection{Study of the VHE $\gamma$-ray spectrum}
\begin{figure}
    \centering
    \begin{subfigure}[t]{0.45\textwidth}
        \centering
       \includegraphics[width=\linewidth]{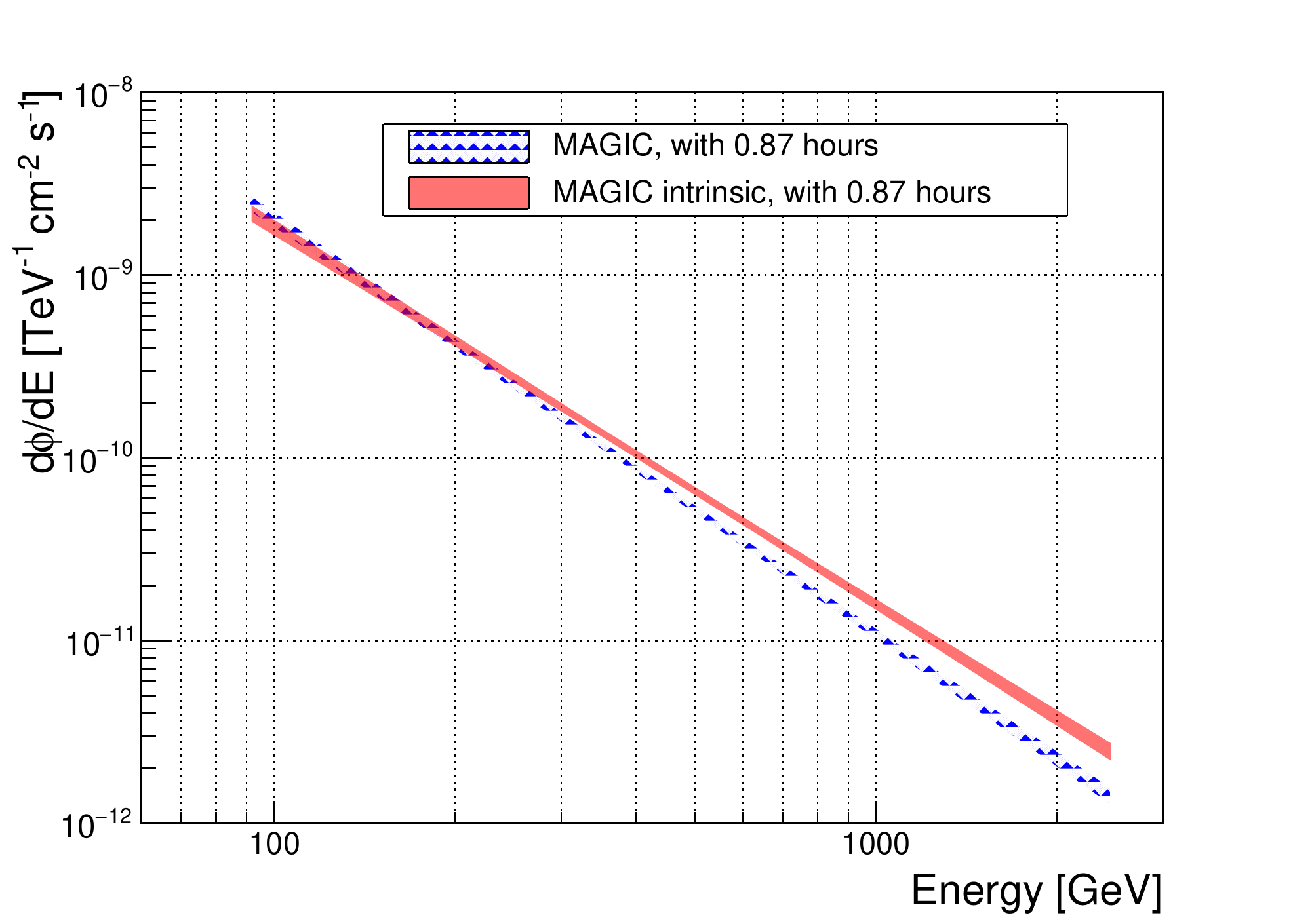} 
        \caption{Observed (blue striped band) and intrinsic (red solid band) spectra. For this Figure the combined spectra from nights MJD~57611 (11 August 2016) and MJD~57612 (12 August 2016) are used.} \label{fig:magic_spectra}
    \end{subfigure}
    \hfill
    \begin{subfigure}[t]{0.45\textwidth}
        \centering
        \includegraphics[width=\linewidth]{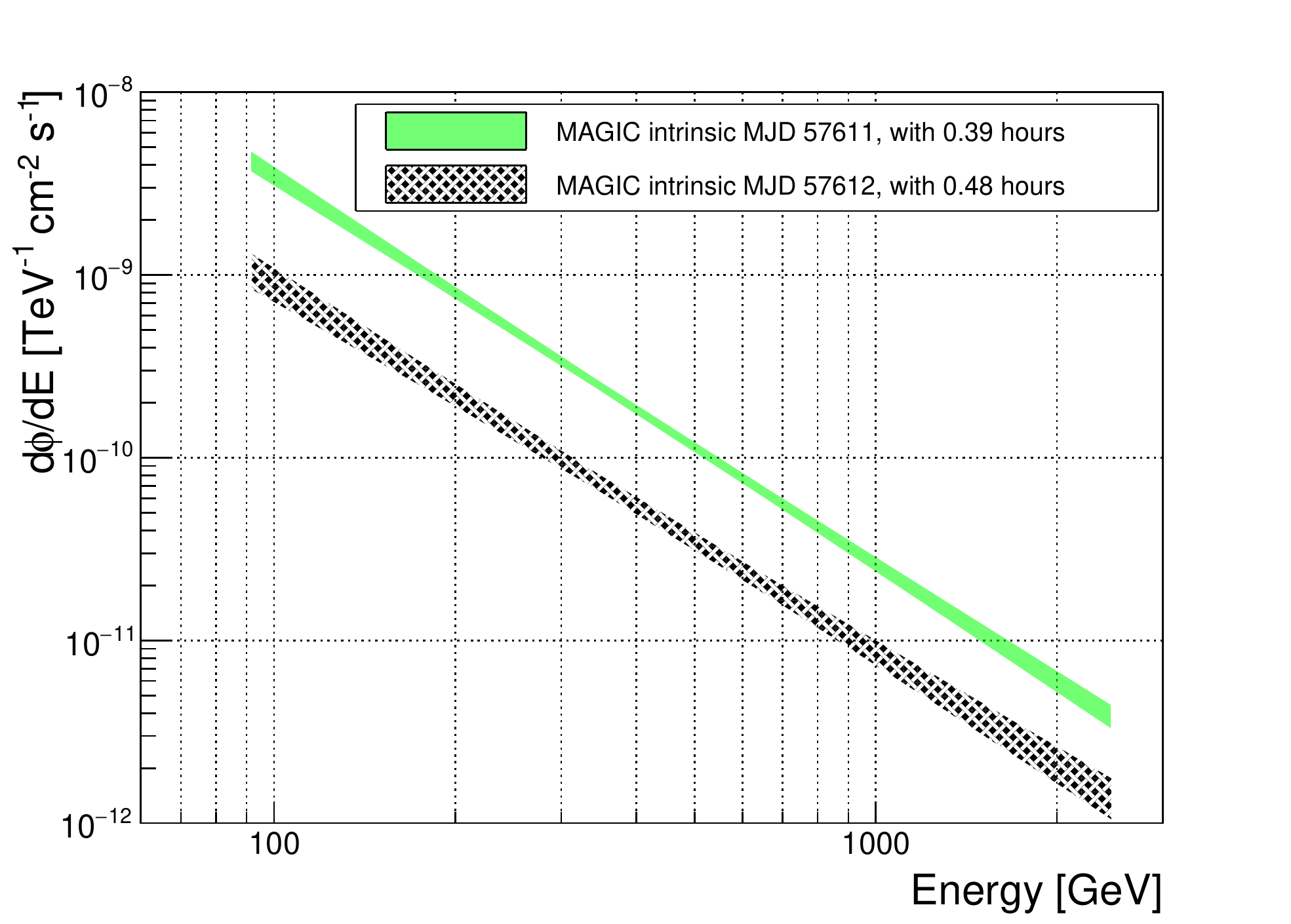} 
        \caption{Intrinsic spectra (corrected for EBL absorption by~\citealt{2011MNRAS.410.2556D}) for MJD~57611 (green solid band) and MJD~57612 (black striped band) respectively.} \label{fig:magic_spectra_MJD}
    \end{subfigure}
  \caption{Differential energy spectra of the VHE $\gamma$-ray emission. The corresponding parameters of the power-law fits are listed in Table~\ref{table:VHEspectrum}.}
\label{fig:magic_all_spectra}
\end{figure}

We gather together all the VHE $\gamma$-ray spectra of 1ES~2344+514 present in the literature so far (up to November 2019). The corresponding SEDs, in the form $E^{2} dN/dE = f_{0}\cdot(E/E_{0})^{\Gamma{+2}}$, are plotted in Fig.~\ref{fig:all_spectra}.
The parameters from the respective power-law fit are listed in Table~\ref{table:gamma_previous}.
The combined differential energy spectrum from MAGIC 2016 observations is shown in Fig.~\ref{fig:magic_spectra}: the (blue) striped band represents the unfolded observed spectrum, while the (red) full band the corresponding spectrum corrected for extragalactic background light (EBL) absorption. In the present work, we use the EBL model of \citet{2011MNRAS.410.2556D} for EBL correction. The spectrum after EBL correction is defined as intrinsic. The combined observed SED of this work is also reported in Fig.~\ref{fig:all_spectra} for comparison purposes. The differential VHE $\gamma$-ray spectra can be successfully described by a simple power law between 90\,GeV and 2.5\,TeV,
\begin{center}
  \begin{equation}
  \centering
    \frac{dF}{dE}=f_0 \left(\frac{E}{500\,~\mathrm{GeV}}\right)^{-\Gamma},
    \label{observed_spectrum}
  \end{equation}
\end{center}
where the normalization constant $f_0$, the spectral index $\Gamma$ and the goodness of the fit ($\chi^2$/d.o.f.) are reported in Table~\ref{table:VHEspectrum}. The best-fit value of the $\Gamma$ index obtained from the combined MAGIC spectrum  is \mbox{$\Gamma = 2.25 \pm 0.12$}. 
The intrinsic spectrum is best described by \mbox{$\Gamma = 2.04 \pm 0.12$}.\par
In Table~\ref{table:gamma_previous} the other values of $\Gamma$ from previous publications in the VHE $\gamma$-ray band are presented. Observed and intrinsic spectra are compared. The $\Gamma$-index values are accompanied by the corresponding flux in percent of the flux of the Crab Nebula. Table~\ref{table:gamma_previous} also shows the energy above which the integral flux is calculated for the corresponding observations. It is not possible to see a clear harder-when-brighter behaviour as typical of HBL. 
Table~\ref{table:VHEspectrum} reveals that the $\Gamma$ indices from MAGIC observations are describing a quite hard spectrum, which maintains its hardness even during the second night of MAGIC observations (MJD~57612 -- 12 August 2019), when the flux drops from the 55\% to 16\% of the Crab Nebula flux above 300\,GeV.
The shift in differential flux between MJD~57611 (11 August 2016) spectrum and the one from MJD~57612 (12 August 2016) is shown in Fig.~\ref{fig:magic_spectra_MJD}. 
Even if a clear decrease in flux is observed between the two consecutive nights, the hardness of the $\Gamma$ index does not vary significantly. Therefore, we do not observe a harder-when-brighter behaviour on short timescales and the position of the IC peak is not changing over the two nights of observation. 
Interestingly, a VHE $\gamma$-ray spectrum with similar hardness was also observed when the source was in a lower activity state (as in \citealt{2013A&A...556A..67A}, $\Gamma$=2.2 when flux is 2.5\% of the Crab Nebula --see Table~\ref{table:gamma_previous}).\par
The MAGIC spectra were also fitted with a log-parabola spectral shape defined by:
\begin{center}
  \begin{equation}
  \centering
    \frac{dF}{dE}=f_0 \left(\frac{E}{500\,~\mathrm{GeV}}\right)^{-\alpha-\beta \log{ \left( \frac{E}{500\,~\mathrm{GeV}} \right). } }
    \label{observed_spectrum_log_parabola}
  \end{equation}
\end{center}

\noindent
The obtained values of the spectral parameters $f_0$, $\alpha$ and $\beta$ are listed in Table~\ref{table:VHEspectrumlogparabola}. No spectral variability is observed and a log-parabola shape is not significantly preferred with respect to a simple power-law. Based on the best-fitted $\alpha$ and $\beta$, the IC peak is located at $\sim 400$\,GeV.\par
As discussed in Sec.~\ref{sec:spectral_Xray}, we confirm that an elevated flux of 1ES~2344+514 can lead to an extreme X-ray state. In the case of Mrk~501 in 2012, \citet{2018A&A...620A.181A} showed that such extreme X-ray states also led to a large shift of the IC peak, to about $1$\,TeV, with hard power-law slopes well below 2 at VHE. Differently, the strong shift of the synchrotron peak of 1ES~2344+514 in 2016 does not seem to be accompanied with a comparable extreme VHE state. The IC peak remains below $1$\,TeV as revealed by the spectral fits. An extreme state in the X-ray regime thus does not always coincide with hard-TeV spectra. In this sense, the 2016 flare of 1ES~2344+514 resembles more the Mrk~501 flare of 1997 \citep{1998ApJ...492L..17P}. This complex EHBL phenomenology is described extensively in recent works~\citep{2019MNRAS.486.1741F,2018MNRAS.477.4257C,2020ApJS..247...16A}.

\section{Broadband SED and modeling}
\label{sec:SED}

\subsection{MWL data and simultaneity}
In Fig.~\ref{fig:ssc_model} we show the broadband SED from radio to VHE energies using quasi-simultaneous observations (red solid squares) around the first \textit{Swift}-XRT observation that took place on MJD~57613 (13 August 2016). For comparison purposes, we add archival data (in dark grey color, small solid dots) taken from the SSDC (ASI Science Data Centre) database\footnote{\url{http://www.asdc.asi.it/}}. \par
For the VHE $\gamma$-ray band, we used the MAGIC SED from the night MJD~57612 (12 August 2016). On that night, the flux in VHE $\gamma$ rays was lower with respect to the first night observed with MAGIC, but closer in time to the first \textit{Swift}-XRT pointing. As shown in Fig.~\ref{fig:magic_spectra_MJD}, despite the difference in flux between the two observations, the spectral slope is in both cases best fitted by an intrinsic index of $\Gamma \approx 2$.\par
From the FACT daily binned light curve (see Fig.~\ref{fig:appendix} of the Appendix~\ref{sec:appendix}), we observe a very similar behaviour in the VHE $\gamma$-ray band between the night MJD~57612 (12 August 2016) for which we have MAGIC data, and the next night MJD~57613 (13 August 2016), for which we do not have MAGIC observations. The measurements by FACT are indeed consistent with a constant flux of $\sim 20$\% that of the Crab Nebula flux above 810\,GeV. Furthermore, we note an absence of spectral variability both in the VHE and X-ray over the MWL campaign (see Fig.~\ref{fig:x_ray_flux_vs_index} and Fig.~\ref{fig:magic_spectra_MJD}). Hence, it implies only a limited bias in terms of flux and spectral shape that is induced in our SED study when assuming the MAGIC spectrum from MJD~57612 (12 August 2016) to be simultaneous with the one measured by \textit{Swift}-XRT on  MJD~57613 (12 August 2016).\par 
We complement the MAGIC data with contemporaneous HE observations provided by the \textit{Fermi}-LAT instrument. The SED points in Fig.~\ref{fig:ssc_model} are integrated over 1 month (centred around the MAGIC observing window, from MJD 57596.5 to MJD 57626.5) due to the faintness of the source for the LAT detector. Despite the low flux, which is a common characteristic among HBL and EHBL, the LAT significantly detected the source over this 1 month period with a TS = 41. The best fit spectral index is \mbox{$\Gamma$ = $1.7\pm0.2$}, while the average over the entire studied period is \mbox{$\Gamma$ = $1.9\pm0.1$}. As already mentioned in Sect.~\ref{sec:MWL}, no strong hint of spectral or flux variability on a weekly timescale was detected in the \textit{Fermi}-LAT light curve. This is consistent with the modest variability of HBL and EHBL in the HE band reported in previous studies \citep{2019MNRAS.486.1741F}. This allows us to assume that the LAT SED points are a good approximation of a LAT SED strictly simultaneous with the MAGIC observations. Additionally, we note a very smooth connection between the \textit{Fermi}-LAT and MAGIC SED.\par

\subsection{Leptonic model description}
\label{sec:leptonic}
We adopt a one-zone SSC model assuming a stationary population of electrons as a first possible emission scenario to describe the broadband SED \citep{2003ApJ...593..667M,2004ApJ...601..151K}. This simple model was already applied to 1ES~2344+514 \citep{2007ApJ...662..892A,2011ApJ...738..169A,2013A&A...556A..67A,2010MNRAS.401.1570T}.\par
For this study we make the following assumptions: 
\begin{itemize}
    \item A spherical homogeneous emission zone with radius $R$ that is filled with relativistic electrons and moving relativistically along the jet with a bulk Lorentz factor $\Gamma_{\rm b}$.
    \item The jet axis is aligned with the line of sight with an angle \mbox{$\Theta$ = 1/$\Gamma_{\rm b}$}; the advantage of such a standard configuration is to reduce the number of degrees of freedom, as the Doppler factor becomes equal to the bulk Lorentz factor (i.e., \mbox{$\delta$ = $\Gamma_{\rm b}$}).
    \item The emitting zone is embedded in an homogeneous magnetic field $B$.
\end{itemize}
We use the open-source software \texttt{naima}~\citep{naima} to compute the synchrotron and IC emissivities.\par
The electron energy distribution (EED) is described by a simple power-law function,
\begin{equation}
    N(\gamma)= N_0\, \gamma^{-n}, \quad \gamma_{\rm min}<\gamma<\gamma_{\rm max},
\end{equation}
where $N_0$ is a normalization constant which is adjusted to match an EED with an energy density of $W_{\rm e}$\,erg\,cm$^{-3}$. The adimensional parameters $\gamma_{\rm min}$ and $\gamma_{\rm max}$ are the minimum and maximum electron Lorentz factors, respectively. This simple parametrization of the EED reduces the number of degrees of freedom and is able to well describe the observed SED.\par
\begin{figure}
	\includegraphics[width=1.1\columnwidth]{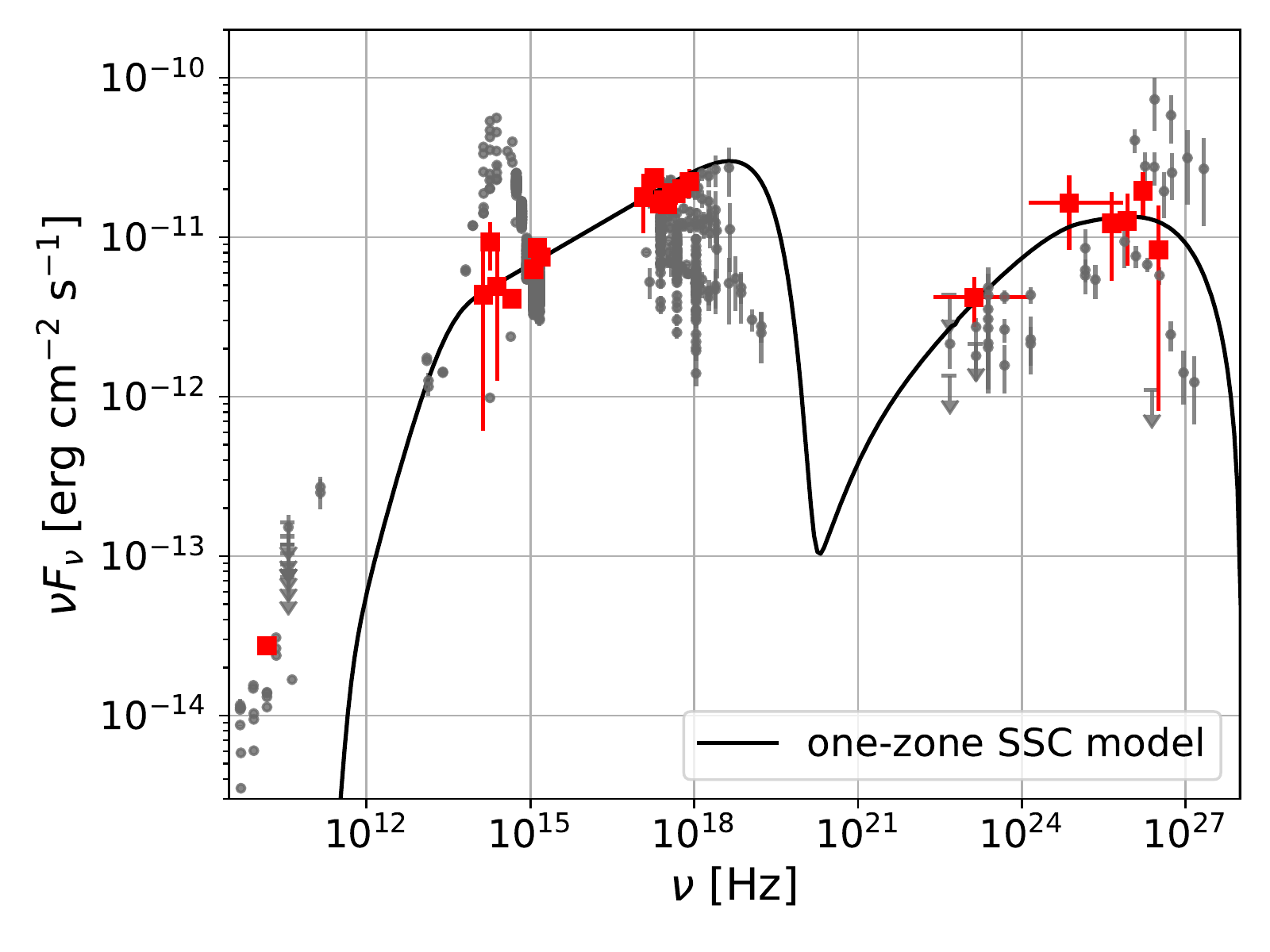}
    \caption{Broadband SED of 1ES~2344+514. Red solid squares represent the simultaneous MWL data used for the modeling. The black solid line is the resulting one-zone SSC model. The VHE $\gamma$-ray SED  shown in the present figure is the intrinsic one.
    Archival data from SSDC are shown by grey solid dots. The shift of the synchrotron peak is clearly visible when comparing the data and modeling of this work to the archival data from SSDC.}
    \label{fig:ssc_model}
\end{figure}

In Table~\ref{table:sscparam}, we list the obtained one-zone SSC parameters, and the corresponding model is plotted in black in Fig.~\ref{fig:ssc_model}. We select a value of $\delta=30$ for the Doppler factor which is typical for HBLs \citep{2010MNRAS.401.1570T}. The size of the emitting region can usually be constrained by the light crossing time \mbox{$R \leq \delta \, t_{\rm var} \, c / \, (1+z)$}, where $t_{\rm var}$ is the observed flux variability timescale. The strong flux decay observed by MAGIC between MJD~57611 and MJD~57612 (11 and 12 August 2016) indicates a day timescale variability at VHE and we therefore set $R=10^{16}$\,cm consistently with the light-crossing time for $\delta=30$. This value is also very similar to the one used in previous models performed on 1ES~2344+514~\citep[see][]{2013A&A...556A..67A,2010MNRAS.401.1570T}.\par
\begin{table}
\caption{One-zone SSC model parameters$^a$.}
\centering          
\begin{tabular}{ l c  }      
\hline\hline 
Parameter & Value \\  % table heading       
\hline
\hline   
$B$ [G]  &  0.02 \\ 
$\delta$ &  30 \\ 
$R$ [$10^{16}$\,cm] & 1\\
$W_{\rm e}$  [erg\,cm$^{-3}$] & 0.008 \\
$n$ & 2.6\\
$\gamma_{\rm min}$ [$10^3$] & 3.0\\
$\gamma_{\rm max}$ [$10^6$] & 3.0\\
\hline 
$^a${See text for a detailed description of each parameter.}
\end{tabular}
\label{table:sscparam}      
\end{table}
In general, we find a very good agreement between the model and the data from infrared to TeV energies. Additionally, the model reproduces well the hardness measured by MAGIC in the VHE $\gamma$ rays ($\Gamma \approx 2$). Only the OVRO data (15\,GHz), as well as all the archival radio data, are not well reproduced and strongly underestimated by the model. The size of the emitting blob (which is constrained by the daily variability timescale at VHE) is small enough such that the radio emission is self-absorbed. Hence, the radio flux likely originates from broader regions of the jet that become transparent at those energies. These regions can have very complex environments and morphologies, which are not included in our model.\par 
The one-zone SSC model in Fig.~\ref{fig:ssc_model} implies $\nu_{\rm s} \approx 4.3 \times 10^{18}$\,Hz $\approx 18$\,keV, similarly to \citet{2000MNRAS.317..743G} where the authors conclude that on MJD~50424 (7 December 1996) the $\nu_{\rm s}$ shifted to energies $\geq 10$\,keV. It corresponds to an impressive shift of the peak to higher energies by roughly two orders of magnitude with respect to the low state reported in \citet{2013A&A...556A..67A}, or from the estimation done by \citet{2018A&A...620A.185N} using archival data.\par
From the model, the peak of the IC component $\nu_{\rm IC}$ is located at \mbox{$\nu_{\rm IC} \approx 9.3 \times 10^{25}$\,Hz $\approx 0.4$\,TeV}. This agrees well with log-parabolic fits discussed in Sec.~\ref{sec:spectral}. The obtained $\nu_{\rm IC}$ is higher by almost 1 order of magnitude with respect to the modeling performed during low states discussed in \citet{2013A&A...556A..67A} and \citet{2011ApJ...738..169A}.\par

\subsection{Hadronic model description}
\label{sec:hadronic}
In general, the low-energy hump of the SED is explained in both leptonic and hadronic models by synchrotron radiation of relativistic electrons. The models differ in the origin of the high energy hump, by IC or associated with the emission by relativistic protons in the jet, respectively. \citet{2000NewA....5..377A} and \cite{2001APh....15..121M} initially explored proton-synchrotron scenarios and showed that they are viable solutions to the $\gamma$-ray emission. One of the major weaknesses of blazar hadronic models is that they often require a very high (super-Eddington) luminosity of the proton population needed to reproduce the observations. This is particularly true for the hadronic modeling of bright FSRQ, as discussed by \citet{2015MNRAS.450L..21Z} and others. For low-luminosity HBL, a successful hadronic modeling can be achieved with total powers well below the Eddington luminosity of the super-massive black hole that powers the jet. We investigate a standard proton-synchrotron scenario using the code described in \citet{2015MNRAS.448..910C}. To limit the number of free parameters of the model, we make the following physical assumptions:

\begin{itemize}
   \item The Doppler factor is fixed to $\delta=30$, a value typical for blazars, and identical to the one used for the leptonic model.
   \item The radius $R$ of the emitting region is constrained by the observed variability timescale $\tau_{\rm var}$ via the usual causality argument as \mbox{$R \leq \delta \textrm{c}\tau_{\rm var}/(1+z)$}, where $\tau_{\rm var}$ has been fixed to one day as described in Sec.~\ref{sec:leptonic}.
   \item The maximum Lorentz factor of protons $\gamma_{\rm p, max}$ is computed by equating the acceleration and cooling timescales: the first one is expressed as \mbox{$\tau_{\rm acc} = (m_{\rm p} {\rm c}/\eta {\rm e} B) \gamma_{\rm p}$}, where $\eta$ is a parameter defining the efficiency of the acceleration mechanism, fixed to $0.1$; the cooling timescales considered here are the adiabatic one, \mbox{$\tau_{\rm ad} \approx R/{\rm c}$}, and the synchrotron one.   
    \item Protons and electrons are supposed to share the same acceleration mechanism, and thus the power-law index of the injected particle distribution is the same: $\alpha_{\rm e,1} = \alpha_{\rm p,1}$
    \item The energy distribution of electrons at equilibrium is computed assuming that the main cooling mechanism is synchrotron radiation, which is always the case for proton-synchrotron solutions characterized by magnetic field values of the order of 10\,G--100\,G.
\end{itemize}

As discussed in \citet{2015MNRAS.448..910C}, the synchrotron radiation by protons, when $\gamma_{\rm p, max}$ is defined via the equation of acceleration and cooling timescales, is characterized by a degeneracy in the $B$--$R$ plane: spectra with the same peak frequency lie on a line in the Log($B$)--Log($R$) plane described by \mbox{$B\propto R^{-2/3}$}. The maximum proton-synchrotron peak frequency $\nu_{\rm s}^{\rm p}$ is defined by the transition from the adiabatic-dominated regime to the synchrotron-cooling dominated regime, and it is equal to \mbox{$1.28\times10^{26} \frac{1}{(1+z)} \frac{(3-\alpha_{\rm p,1})}{1.5} \frac{\delta}{10} \mathrm{Hz}$}.\par
We systematically study the parameter space scanning over $\nu_{\rm s}^{\rm p}$, $R$ and on the normalization of the proton distribution $K_{\rm p}$. We produce 1500 models in the following parameter space:  $\nu_{\rm s}^{\rm p} \in [0.1\nu_{\rm max}^{\rm p}, \nu_{\rm max}^{\rm p}]$, $R\in [10^{14} \textrm{cm}, R_{\rm max}]$, and the proton normalisation $K_{\rm p} \in [K^\star/3, 3K^\star]$, where $K^{\star}$ corresponds to the proton density which provides a synchrotron spectrum at the level of the MAGIC spectra. We compute \textit{a posteriori} the $\chi^{2}$ of all models with respect to the data, identify the solution with the minimum $\chi^{2}$, and select only solutions which are comprised within a $\Delta\chi^{2}$ corresponding to $1\sigma$.\par
\begin{figure}
	\includegraphics[width=1.1\columnwidth]{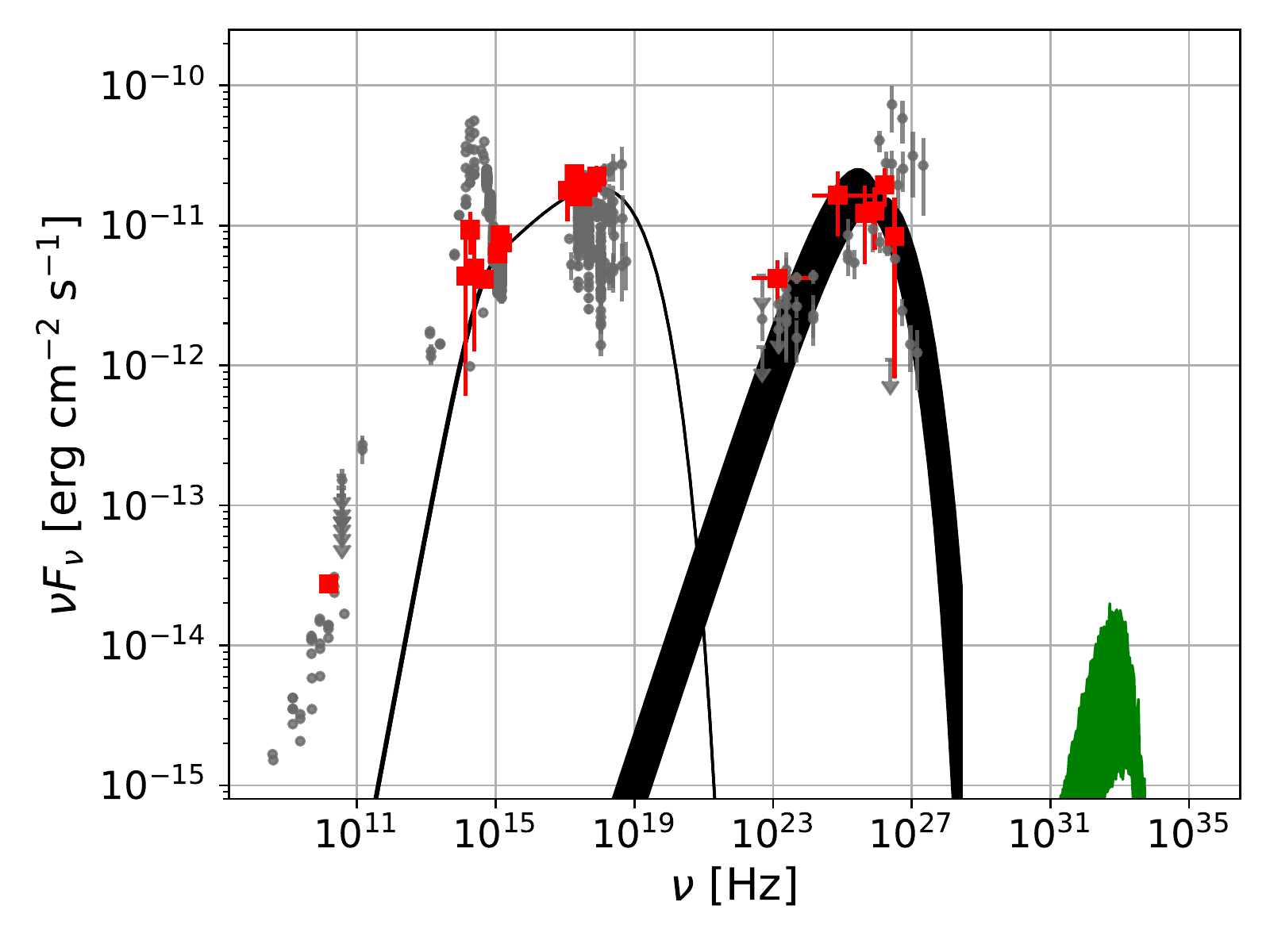}
    \caption{Broadband SED of 1ES~2344+514. Red solid squares represent the simultaneous MWL data used for the modeling. The black band represents the resulting set of hadronic models, while the green band show the corresponding expected neutrino flux, which is lower than IceCube sensitivity for point-like sources \citep[$4.8\times 10^{-13}\mathrm{erg}\,\mathrm{cm}^{-2} \mathrm{s}^{-1} $,][]{2019EPJC...79..234A}. The VHE $\gamma$-ray SED  shown in the present figure is the intrinsic one.
    Archival data from SSDC are shown by grey solid dots.}
    \label{fig:psynch_model}
\end{figure}

The optical and NIR data are compatible with a spectral break associated with the synchrotron self-absorption located around this energy band. This feature breaks the degeneracy in the $B$--$R$ plane, resulting in a rather well constrained value for the magnetic field, the emitting region size, as well as the maximum energy of the proton distribution (see Table~\ref{tab:hadronic-model}). The resulting set of models are plotted as a black band in Fig.~\ref{fig:psynch_model}. In addition, we show with a green band the expected flux of neutrino arising from p-$\gamma$ interactions. In our particular case, the latter component is rather low for the following reason: the particle density for both protons and leptons (which are producing the target photon field for p-$\gamma$ interactions) is much lower compared to the leptonic case in order to compensate for the large magnetic field (\mbox{$B \approx 50$\,G}) and also for suppressing the SSC contribution and let the proton-synchrotron dominate. \par
Overall, there is a rather good agreement between the models and the data points from optical to TeV. The high energy hump is narrower compared to the leptonic model and this creates a small tension with the spectral shapes as seen by \textit{Fermi}-LAT and MAGIC. The models tend to give a harder HE $\gamma$-ray spectrum than the one measured by \textit{Fermi}-LAT, while the opposite trend is visible regarding the MAGIC data. 
From this point of view, the high energy hump is better described in the leptonic model. We note that the shape of the high-energy hump is directly linked to the spectral slope of the proton energy distribution. The latter is obtained based on the assumptions that electrons and protons are accelerated following the same mechanisms (i.e., $\alpha_{\rm p,1} = \alpha_{\rm e,1}$). Since we further assume the synchrotron radiation to be the main cooling mechanism for the electrons, this implies \mbox{$\alpha_{\rm p,1} = \alpha_{\rm e,1}=1.5$}, in order to have \mbox{$\alpha_{\rm e,2} = 2.5$}. Thus, relaxing one of these assumptions will result in a better description of the high energy component.\par

\begin{table}
\caption{Parameters of the hadronic model$^a$.}
    \centering
   		\begin{tabular}{@{}l c}
 		\hline
 		\hline
		 & Proton-synchrotron\\
 		\hline
 		 \noalign{\smallskip}
 		$\delta$ & \textit{30} \\
 $R$ [10$^{15}$ cm] & $0.9-1.9$ \\
 $^\star \tau_{\rm obs}$ [days] & $0.3-0.6$ \\
 		\hline
 		 $B$ [G] & $48-63$   \\
 		$^\star u_B$ [erg cm$^{-3}$] & $94-157$ \\
 		\hline
 		$\gamma_{\rm e,min} $& $200$   \\
 		$\gamma_{\rm e,break} $& $=\gamma_{\rm e,min}$     \\
 		$\gamma_{\rm e,max}\ [10^4]$& $2.6-3.0$\\
 		$\alpha_{\rm e,1}=\alpha_{p,1}$ & $1.5$  \\
 		$\alpha_{\rm e,2}=\alpha_{p,2}$ & $2.5$   \\
 		$K_{\rm e}$ [cm$^{-3}$] & $0.2-1.6$    \\
 		$^\star u_{\rm e}$ [10$^{-5}\,$erg$\,$cm$^{-3}$] & $0.3-3.3$\\
 		\hline
 		$\gamma_{\rm p,min}$& 1  \\
 		$\gamma_{\rm p,break} [10^9]$&  $=\gamma_{\rm p,max}$   \\
 		$\gamma_{\rm p,max} [10^9]$& $3.1-6.8$\\
         $\eta$ [10$^{-4}$]& $2.5-12.9$ \\
 		$^\star u_{\rm p}$ [10$^{-2}\,$erg cm$^{-3}$] &$0.9-30.5$  \\
 		\hline
    	$^\star u_{\rm p}/u_{\rm B} $ [10$^{-4}$]& $0.7-26.3$ \\
 		$^\star L$ [10$^{45}$ erg s$^{-1}$] & $4.1-19.3$  \\
 		\hline
 		\hline
 		\end{tabular}
 	 \newline
 	 $^a${The quantities flagged with a star ($^\star$) are derived quantities, and not model parameters. The luminosity of the emitting region has been calculated as \mbox{$L=2 \pi R^2c\Gamma_{\rm bulk}^2(u_{\rm B}+u_{\rm e}+u_{\rm p})$}, where \mbox{$\Gamma_{\rm bulk}=\delta/2$}, and $u_{\rm B}$, $u_{\rm e}$, and $u_{\rm p}$ are (respectively) the energy densities of the magnetic field, the electrons, and the protons.}
 		      	 \label{tab:hadronic-model}
 		\end{table}

\section{Discussion of the extreme behaviour}
\label{sec:extreme}
The MWL observations presented in this paper clearly reveal a renewed extreme behaviour accompanied with a particularly high state in the VHE range during August 2016. On MJD~57613 (13 August 2016), the spectral analysis in the range 0.3\,keV--10\,keV results in a power-law index harder than 2, with a flux as high as during the December 1996 flare. This strongly suggests a location of the synchrotron peak \mbox{$\gtrsim 10^{18}$ Hz}, thus fulfilling the criteria to be classified as ``extreme'' following \citet{2001A&A...371..512C}. 
Recent published works proposed that there is not a single population of EHBL: some of them show very soft VHE spectra, while others have hard-VHE spectra \citep{2019MNRAS.486.1741F,2018MNRAS.477.4257C}. The VHE $\gamma$-ray spectrum measured here by MAGIC (spectral index of $\sim 2$) possibly indicates 1ES~2344+514 as a transitional source between HBL-like EHBL and hard-TeV EHBL. The latter class of HBL is characterized by spectral indices significantly harder than 2~\citep{2019MNRAS.486.1741F}.\par  
Within the SSC scenario, a shift of the complete broadband SED is confirmed. Based on the model parameters, we derive a \mbox{$\nu_{\rm s} \approx 4.3\times10^{18}$\,Hz $\approx 18$\,keV} and a \mbox{$\nu_{\rm IC} \approx 9.3 \times 10^{25}$\,Hz $\approx 0.4$\,TeV}.
These are significantly different from the results obtained during low emission states. Using the archival data, \citet{2018A&A...620A.185N} estimated a $\nu_{\rm s}$ at $2.2 \times 10^{16}$ Hz.\par 
The impressive displacements of both the synchrotron and IC peaks imply an unusual increase of the energy of the emitting electrons. Accordingly, we find that $\gamma_{\rm max}$ must reach values of a few times \mbox{$10^{6}$}, and thus indicate that the EED extends without significant cutoff or break up to TeV energies. 
In all previous works, during low states, the SSC modeling resulted in a EED showing a clear break at values around \mbox{$10^{4}$--$10^{5}$}, which is about 1 order of magnitude lower than the $\gamma_{\rm max}$ found here. \citet{2015MNRAS.451..611B} similarly inferred that the EED extend without clear break to $\sim 10^{6}$ to model a small sample of EHBL.\par 
Additionally to the extreme value of $\gamma_{\rm max}$, we find that a low magnetic field $B \approx 0.02$\,G is necessary to describe the SED. The magnetic field usually lies between 0.1-1\,G in leptonic models for HBL \citep[see][]{2010MNRAS.401.1570T}, but can be as low as \mbox{$10^{-2}$--$10^{-3}$\,G} in the case of EHBL. A low magnetic field is required to account for an IC component peaking close to the TeV regime as well as the large separation from $\nu_{\rm s}$. In SSC models, we expect a dependency between the magnetic field (and Doppler factor) and the peaks of the two emission bumps as \mbox{$B/\delta \propto \nu_{\rm s}/\nu_{\rm IC}^2$}~\citep{2016MNRAS.456.2374T}.\par
As a direct consequence of a low magnetic field combined with an EED extending up to a few TeV, we obtain an emitting region which is far below energy equipartition, i.e. \mbox{$U_{\rm B}/U_{\rm e} \ll 1$}, where $U_{\rm B}$ is the magnetic energy density and $U_{\rm e}$ is the electrons energy density. Based on the resulting SSC parameters and calculating the jet energetics following the prescription in \citet{2008MNRAS.385..283C}, we get \mbox{$U_{\rm B}/U_{\rm e} \approx 2 \times 10^{-3}$}. Hence, the magnetic energy density is extremely low compared to the energy density stored in the EED. This feature is commonly seen in SED modeling of BL~Lac objects during a flaring episode, such as the one we observed in this work.
Therefore, it sets challenges to explain the electron acceleration mechanism, which is generally thought to be done through the transfer of the magnetic energy to kinetic energy, up to the point where equipartition is reached \citep{2007MNRAS.380...51K}. Interestingly, \citet{2018MNRAS.477.4257C} modeled a small set of EHBL and showed that an energy equipartition far below unity can also persists during quiescent states. We also computed $U_{\rm B}/U_{\rm e}$ based on the models parameters obtained in previous works on 1ES~2344+51.4 \citep{2013A&A...556A..67A,2011ApJ...738..169A,2010MNRAS.401.1570T,2007ApJ...662..892A}. All of them result in values of typically \mbox{$U_{\rm B}/U_{\rm e} \approx 10^{-2}$--$10^{-3}$}.\par 
A similar consideration is made in \cite{acciari2019EHBL}, where a catalog of EHBL is presented and studied in a MWL context using three different modelings. While all modelings give a good description of the observations, the obtained magnetisations substantially differ. A single-zone SSC model applied to their data requires a critically low magnetization, consistent with the results shown in this paper. The proton-synchrotron model was instead providing a highly magnetized jet, still far from equipartition. On the other hand, adopting the so-called spine-layer model \citep{2005A&A...432..401G}, a two-component SSC model comprising a structured jet as emission zone, a quasi-equipartition of the magnetic field and matter could be achieved. Nevertheless, with the data at hand, no model was favoured.\par 
The contemporaneous \textit{Fermi}-LAT spectral points were obtained based on a significant detection of the source on a 1-month timescale. Up to now, no significant spectral points in the HE band have been combined with a contemporaneous VHE spectrum obtained on such a short timescale for 1ES~2344+514. The combined MAGIC and \textit{Fermi}-LAT SED is better described when $\gamma_{\rm min}$ is around \mbox{$3 \times 10^{3}$} rather than close to unity. The main reason for this is that, when reducing $\gamma_{\rm min}$, we increase the pool of electrons that are dominantly responsible for the rising edge of the IC bump. Consequently, the IC flux increases in the HE band. A high $\gamma_{\rm min}$ around \mbox{$3 \times 10^{3}$} therefore provides a narrower IC peak. A high minimum energy in the EED is a recurrent feature in EHBL. We note that $\gamma_{\rm min}$ is constrained by the (host-galaxy corrected) IR/optical data and can not be increased arbitrarily high.\par
The proton-synchrotron models also describe the data well, with a electrons synchrotron peak frequency at \mbox{${\sim}9\times10^{17}$\,Hz.}\par
With respect to the leptonic model, the hadronic models require a much larger magnetic field indicating that the emission zone has an equipartition parameter well above 1, generating the opposite situation than in the SSC model. As seen in Tab.~\ref{tab:hadronic-model}, the magnetic field energy densities are $10^{2}$--$10^{4}$ higher than the energy density of the particles in the jet.\par
 The low luminosity commonly found in EHBL does not require super-Eddington luminosity for the proton population and we find a maximum total jet luminosity of \mbox{$\sim 2\times10^{45}$\,erg\,s$^{-1}$}. For supermassive black holes of $\sim10^9$ solar masses, as estimated for 1ES~2344+514 \citep{2003ApJ...583..134B}, Eddington luminosity is about 1 order of magnitude higher.\par 
 For the reasons explained in Sect.~\ref{sec:hadronic}, the predicted neutrino flux resulting from $p-\gamma$ interactions is much lower than the high energy hump of the SED and lies well below the sensitivity of the IceCube neutrino detector. Thus, within pure hadronic scenarios, no neutrino are expected to be detected by current neutrino detectors, even during flaring events of 1ES2344+514. In agreement with our results, the IceCube Collaboration evaluated a 90\% CL upper limit of the muon neutrino flux of 1ES~2344+514  at \mbox{$8.5\times 10^{-13}\mathrm{erg}\,\mathrm{cm}^{-2}\, \mathrm{s}^{-1} $} \citep{2019EPJC...79..234A}.\par
In our hadronic modeling, we obtain that, by construction, the shortest cooling timescale is the adiabatic one, and the proton population shows no cooling break ($\gamma_{\rm p,break}=\gamma_{\rm p,max}$). As shown in Tab.~\ref{tab:hadronic-model}, the variability timescale expected from adiabatic cooling is below or around one day and remains consistent with the daytime scale variability observed in the present flare.\par
The models we applied to the present broadband dataset can both successfully describe the MWL SED.\\

\section{Summary and Conclusions}
\label{sec:summary}
Triggered by the FACT detection of enhanced flux in the TeV range, on MJD~57611 (11 August 2016) the MAGIC telescopes started to observe the BL~Lac object 1ES~2344+514.\par
The MAGIC observations resulted in a detection with a significance of $13\sigma$ in less than one hour and a measured flux of 55\% of the Crab Nebula flux above 300\,GeV. This flux is comparable with the historical maximum detected from this source in 1995 \citep{1998ApJ...501..616C}. On the following night (MJD~57612--12 August 2016) the signal was already fading and the measured flux was 16\% of the Crab Nebula flux above 300\,GeV. We gathered MWL data from instruments in the radio, optical, NIR, UV, X-ray and HE band to complement the VHE $\gamma$-ray observations: with simultaneous data taken on MJD~57613 (13 August 2016) we built a broadband SED describing the flaring state and modeled it within two alternative scenarios: a leptonic SSC, and a proton-synchrotron model. \par
For the first time in this source, \textit{Fermi}-LAT data with MAGIC data allow us to constrain the IC hump on short timescales. A leptonic model applied to the data gives a peak frequency \mbox{$\nu_{\rm IC} \approx 9.3 \times 10^{25}$\,Hz $\approx 0.4$\,TeV.} . \par
We find the source in an extreme synchrotron state, with a peak frequency obtained from the leptonic model at \mbox{$\nu_{\rm s} \approx 4.3\times10^{18}$\,Hz}, corresponding to \mbox{$\sim18$\,keV}. The shift of the $\nu_{\rm s}$ with respect to previous observations~\citep{2013A&A...556A..67A, 2018A&A...620A.185N} is of about two orders of magnitude.\par 
We also find a harder than usual VHE $\gamma$-ray spectrum (\mbox{$\Gamma = 2.04 \pm 0.12_{\rm stat} \pm 0.15_{\rm sys}$}  after EBL correction). The hardness of the spectrum does not vary between the first and the second night of observation, even if the latter one is characterized by a three times lower flux.\par
The leptonic and hadronic models both describe successfully the data. On the other hand, they imply a significantly different magnetization of the emitting zone.\par
We conclude that the BL~Lac object 1ES~2344+514 belongs to that subcategory of EHBL which reveal to be extreme only in some circumstances \citep[see Mrk~501 in 2012;][]{2018A&A...620A.181A}, and does not show the typical characteristic of persistent extreme SED as for instance the archetypal EHBL 1ES~0229+200 does~\citep{2007A&A...475L...9A}. This ``intermittent'' extremeness could be studied acquiring more MWL data in the next years. Time-dependent modelling to interpret the broadband SED could help to elucidate this peculiarity. \par
There is still more to discover about the EHBL family and future MWL campaigns will help to unveil their nature and to move towards a classification of those interesting powerful AGN.

\section*{Acknowledgements}
%
% MAGIC STANDARD ACKNOWLEDGEMENTS
%
We would like to thank the Instituto de Astrof\'{\i}sica de Canarias for the excellent working conditions at the Observatorio del Roque de los Muchachos in La Palma. The financial support of the German BMBF and MPG, the Italian INFN and INAF, the Swiss National Fund SNF, the ERDF under the Spanish MINECO (FPA2017-87859-P, FPA2017-85668-P, FPA2017-82729-C6-2-R, FPA2017-82729-C6-6-R, FPA2017-82729-C6-5-R, AYA2015-71042-P, AYA2016-76012-C3-1-P, ESP2017-87055-C2-2-P, FPA2017-90566-REDC), the Indian Department of Atomic Energy, the Japanese JSPS and MEXT, the Bulgarian Ministry of Education and Science, National RI Roadmap Project DO1-153/28.08.2018 and the Academy of Finland grant nr. 320045 is gratefully acknowledged. This work was also supported by the Spanish Centro de Excelencia ``Severo Ochoa'' SEV-2016-0588 and SEV-2015-0548, and Unidad de Excelencia ``Mar\'{\i}a de Maeztu'' MDM-2014-0369, by the Croatian Science Foundation (HrZZ) Project IP-2016-06-9782 and the University of Rijeka Project 13.12.1.3.02, by the DFG Collaborative Research Centers SFB823/C4 and SFB876/C3, the Polish National Research Centre grant UMO-2016/22/M/ST9/00382 and by the Brazilian MCTIC, CNPq and FAPERJ.
%FACT: 
The FACT collaboration acknowledges the important contributions from ETH Zurich grants ETH-10.08-2 and ETH-27.12-1 as well as the funding by the Swiss SNF and the German BMBF (Verbundforschung Astro- und Astroteilchenphysik) and HAP (Helmoltz Alliance for Astroparticle Physics) are gratefully acknowledged. Part of this work is supported by Deutsche Forschungsgemeinschaft (DFG) within the Collaborative Research Center SFB 876 "Providing Information by Resource-Constrained Analysis", project C3. We are thankful for the very valuable contributions from E. Lorenz, D. Renker and G. Viertel during the early phase of the project. We thank the Instituto de Astrof\'{\i}sica de Canarias for allowing us to operate the telescope at the Observatorio del Roque de los Muchachos in La Palma, the Max-Planck-Institut f\"{u}r Physik for providing us with the mount of the former HEGRA CT3 telescope, and the MAGIC collaboration for their support.
%WEBT
This article is based partly on observations made with the 1.5 TCS and IAC80 telescopes operated by the IAC in the Spanish Observatorio del Teide. This article is also based partly on data obtained with the STELLA robotic telescopes in Tenerife, an AIP facility jointly operated by AIP and IAC. 
%Valeri
We acknowledge support from Russian Scientific Foundation grant 17-12-01029.
%KAIT
A.V.F. and W.Z. are grateful for support from NASA grant NNX12AF12G, the Christopher R. Redlich Fund, the TABASGO Foundation, and the Miller Institute for Basic Research in Science (U.C. Berkeley). KAIT and its ongoing operation were made possible by donations from Sun Microsystems, Inc., the Hewlett-Packard Company, AutoScope Corporation, Lick Observatory, the US National Science Foundation, the University of California, the Sylvia and Jim Katzman Foundation, and the TABASGO Foundation. Research at Lick Observatory is partially supported by a generous gift from Google.\par
W.M. acknowledges support from CONICYT project Basal AFB-170002.
%OVRO
The OVRO 40-m monitoring program is supported in part by NASA grants NNX08AW31G, NNX11A043G, and NNX14AQ89G, and NSF grants AST-0808050 and AST-1109911. 
%Swift
This research has made use of data and/or software provided by the High Energy Astrophysics Science Archive Research Center (HEASARC), which is a service of the Astrophysics Science Division at NASA/GSFC and the High Energy Astrophysics Division of the Smithsonian Astrophysical Observatory. We acknowledge the use of public data from the {\it Swift} data archive.
%teVcat
This research has made use the TeVCat online source catalog (http://tevcat.uchicago.edu). 
%SSdC
Part of this work is based on archival data, software or online services provided by the Space Science Data Center - ASI.

\section*{Affiliations}
\noindent
{\it
$^{1}$ {Inst. de Astrof\'isica de Canarias, E-38200 La Laguna, and Universidad de La Laguna, Dpto. Astrof\'isica, E-38206 La Laguna, Tenerife, Spain} \\
$^{2}$ {Universit\`a di Udine, and INFN Trieste, I-33100 Udine, Italy} \\
$^{3}$ {National Institute for Astrophysics (INAF), I-00136 Rome, Italy} \\
$^{4}$ {ETH Zurich, CH-8093 Zurich, Switzerland} \\
$^{5}$ {Technische Universit\"at Dortmund, D-44221 Dortmund, Germany} \\
$^{6}$ {Croatian Consortium: University of Rijeka, Department of Physics, 51000 Rijeka; University of Split - FESB, 21000 Split; University of Zagreb - FER, 10000 Zagreb; University of Osijek, 31000 Osijek; Rudjer Boskovic Institute, 10000 Zagreb, Croatia} \\
$^{7}$ {Saha Institute of Nuclear Physics, HBNI, 1/AF Bidhannagar, Salt Lake, Sector-1, Kolkata 700064, India} \\
$^{8}$ {Centro Brasileiro de Pesquisas F\'isicas (CBPF), 22290-180 URCA, Rio de Janeiro (RJ), Brasil} \\
$^{9}$ {IPARCOS Institute and EMFTEL Department, Universidad Complutense de Madrid, E-28040 Madrid, Spain} \\
$^{10}$ {University of Lodz, Faculty of Physics and Applied Informatics, Department of Astrophysics, 90-236 Lodz, Poland} \\
$^{11}$ {Universit\`a  di Siena and INFN Pisa, I-53100 Siena, Italy} \\
$^{12}$ {Deutsches Elektronen-Synchrotron (DESY), D-15738 Zeuthen, Germany} \\
$^{13}$ {Istituto Nazionale Fisica Nucleare (INFN), 00044 Frascati (Roma) Italy} \\
$^{14}$ {Max-Planck-Institut f\"ur Physik, D-80805 M\"unchen, Germany} \\
$^{15}$ {Institut de F\'isica d'Altes Energies (IFAE), The Barcelona Institute of Science and Technology (BIST), E-08193 Bellaterra (Barcelona), Spain} \\
$^{16}$ {Universit\`a di Padova and INFN, I-35131 Padova, Italy} \\
$^{17}$ {Universit\`a di Pisa, and INFN Pisa, I-56126 Pisa, Italy} \\
$^{18}$ {Universitat de Barcelona, ICCUB, IEEC-UB, E-08028 Barcelona, Spain} \\
$^{19}$ {The Armenian Consortium: ICRANet-Armenia at NAS RA, A. Alikhanyan National Laboratory} \\
$^{20}$ {Centro de Investigaciones Energ\'eticas, Medioambientales y Tecnol\'ogicas, E-28040 Madrid, Spain} \\
$^{21}$ {Universit\"at W\"urzburg, D-97074 W\"urzburg, Germany} \\
$^{22}$ {Finnish MAGIC Consortium: Finnish Centre of Astronomy with ESO (FINCA), University of Turku, FI-20014 Turku, Finland; Astronomy Research Unit, University of Oulu, FI-90014 Oulu, Finland} \\
$^{23}$ {Departament de F\'isica, and CERES-IEEC, Universitat Aut\`onoma de Barcelona, E-08193 Bellaterra, Spain} \\
$^{24}$ {Japanese MAGIC Consortium: ICRR, The University of Tokyo, 277-8582 Chiba, Japan; Department of Physics, Kyoto University, 606-8502 Kyoto, Japan; Tokai University, 259-1292 Kanagawa, Japan; RIKEN, 351-0198 Saitama, Japan} \\
$^{25}$ {Inst. for Nucl. Research and Nucl. Energy, Bulgarian Academy of Sciences, BG-1784 Sofia, Bulgaria} \\
$^{26}$ {now at University of Innsbruck}\\
$^{27}$ {also at Port d'Informaci\'o Cient\'ifica (PIC) E-08193 Bellaterra (Barcelona) Spain}\\
$^{28}$ {also at Dipartimento di Fisica, Universit\`a di Trieste, I-34127 Trieste, Italy}\\
$^{29}$ {also at INAF-Trieste and Dept. of Physics \& Astronomy, University of Bologna}\\
$^{30}$ {University of Geneva, Department of Astronomy, Chemin d'Écogia 16, 1290 Versoix, Switzerland} \\
$^{31}$ {also at RWTH Aachen University}\\
$^{32}$ {Department of Astronomy, University of California, Berkeley, CA 94720-3411, USA}\\
$^{33}$ {Miller Senior Fellow, Miller Institute for Basic Research in Science, University of California, Berkeley, CA  94720, USA}\\
$^{34}$ {Finnish Centre for Astronomy with ESO (FINCA), University of Turku, FI-20014, Turku, Finland}\\
$^{35}$  {Aalto University Mets\"ahovi Radio Observatory, Mets\"ahovintie 114, 02540 Kylm\"al\"a, Finland}\\
$^{36}$ {Owens Valley Radio Observatory, California Institute of Technology, Pasadena, CA 91125, USA}\\
$^{37}$ {Astronomical Institute, St. Petersburg State University, Universitetskij Pr. 28, Petrodvorets, St. Petersburg 198504, Russia}\\
$^{38}$ {Main (Pulkovo) Astronomical Observatory of RAS, Pulkovskoye shosse 60, St. Petersburg 196149, Russia}\\
$^{39}$ {Departamento de Astronom\'ia, Universidad de Chile, Camino El Observatorio 1515, Las Condes, Santiago, Chile}\\
$^{40}$ {INAF, Osservatorio Astrofisico di Torino, I-10025 Pino Torinese, Italy}} \\
$^{\color{blue}\varheartsuit}$ {currently at $^{3}$ }\\% National Institute for   Astrophysics (INAF),   I-00136 Rome, Italy}\\
%%%%%%%%%%%%%%%%%%%%%%%%%%%%%%%%%%%%%%%%%%%%%%%%%%
\section*{Data availability}
The complete dataset shown in Fig.~\ref{fig:MWL_lc} and Fig.~\ref{fig:magic_all_spectra}, the data points shown in Fig.~\ref{fig:ssc_model} and Fig.~\ref{fig:psynch_model} and Table~\ref{table:VHEspectrum} are  available at the CDS  http://cdsarc.u-strasbg.fr .\\
Other data underlying this article will be shared on reasonable request to the corresponding authors.

%%%%%%%%%%%%%%%%%%%% REFERENCES %%%%%%%%%%%%%%%%%%

% The best way to enter references is to use BibTeX:

\bibliographystyle{mnras}
\bibliography{biblio.bib} 

%%%%%%%%%%%%%%%%%%%%%%%%%%%%%%%%%%%%%%%%%%%%%%%%%%

%%%%%%%%%%%%%%%%% APPENDICES %%%%%%%%%%%%%%%%%%%%%
\appendix
\section{Insights on FACT analysis}
\label{sec:appendix}
The quality of the data is evaluated using an artificial trigger rate having a threshold that is set with a digital-to-analog (DAC) converter in DAC counts \citep{2013JInst...8P6008A}. For this analysis, the artificial trigger rate above a threshold of 750\,DAC-counts (hereafter $R750$) is calculated. This threshold is high enough such that accidental triggers are highly suppressed and the measured $R750$ rate is due to cosmic-ray induced air showers. Evaluating the dependence of $R750$ on the zenith distance, a corrected rate $R750_{\rm cor}$ is calculated. \par
To account for seasonal changes of the cosmic-ray rate due to variations in Earth's atmosphere, a reference value $R750_{\rm ref}$ is determined for each moon period. Data with good quality are selected using a cut of $R750_{\rm cor}/R750_{\rm ref} > 0.7$. This rather conservative cut was chosen, as part of the data (including the flare) were taken, when the weather phenomenon Calima occurred, a.k.a.\ Saharan Air Layer (SAL), i.e.\ a layer carrying dust from the Sahara which can extend from the African coast to the Caribbean~\citep{2009A&A...493..721D,2014arXiv1403.3591F}. \par
\begin{figure}

	\includegraphics[width=1.1\columnwidth]{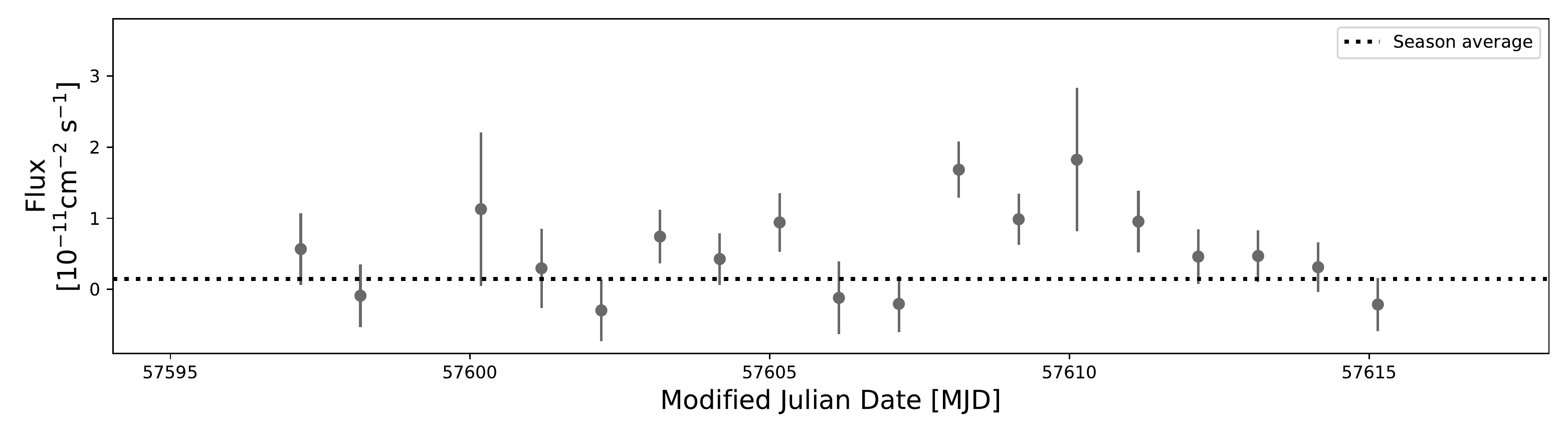}
    \caption{Daily light curve obtained with FACT from the time period MJD~57595 to MJD~57620 (26 July to 20 August 2016)
    }
    \label{fig:appendix}
\end{figure}

As the SAL absorbs Cherenkov photons, the observed size of the showers is reduced. Consequently, the reconstructed energy of $\gamma$-ray showers, which is mainly proportional to the size of the shower, is biased (i.e., the observed energy is lower than the true energy of the incoming $\gamma$ ray) and the trigger efficiency decreases resulting in a reduction of the reconstructed $\gamma$-ray flux. This also affects the cosmic-ray rate and its dependency on the dust concentration is shown in \cite{2019ICRC...36..630B}. Therefore, the standard data selection cut of $0.93 < R750_{\rm cor}/R750_{\rm ref} < 1.3$ would have cut away a large fraction the data suffering from the SAL. Instead, we adopt the above-mentioned conservative cut and apply a correction to the $\gamma$-ray flux. 
Given its strong dependency on the SAL, the $R750_{\rm cor}$ can be used to estimate the energy bias in the observed $\gamma$-ray showers. First, under the assumption of a constant cosmic-ray flux (following a power law with index $-2.7$), the reduction in the $R750_{\rm cor}$ can be translated into a bias in the observed energy that is responsible for the measured decrease in the cosmic-ray rate. Secondly, assuming that the portion of Cherenkov light affected by the SAL is similar between hadronic and $\gamma$-ray showers, the estimated energy bias is further used to calculate a correction factor applied to the measured $\gamma$-ray flux. 
We consider here that 1ES 2344+514 follows a power-law with index $-2.46$ (\cite{2017MNRAS.471.2117A}). The correction factors are calculated on a nightly basis and consist of at most $\sim 30$\% percent of the flux, thus remaining within the statistical uncertainties. 
We complement the FACT observations reported in Sec.~\ref{sec:MWL} with the daily light curve around the flare in Fig.~\ref{fig:appendix}.

%%%%%%%%%%%%%%%%%%%%%%%%%%%%%%%%%%%%%%%%%%%%%%%%%%

% Don't change these lines
\bsp	% typesetting comment
\label{lastpage}
\end{document}